\documentclass[aps,prx,amsmath,amssymb,lengthcheck,showpacs,floatfix,groupedaddress,superscriptaddress,longbibliography]{revtex4-2}
\usepackage{placeins}
\usepackage{array,longtable}
\usepackage{tabularray}
\usepackage{enumitem}
\usepackage{siunitx,soul}
\usepackage{color,makecell,adjustbox,multirow}
\usepackage{hhline}
\usepackage{amsmath}
\usepackage{amsfonts}
\usepackage{amssymb}
\usepackage{graphicx}
\usepackage[english]{babel}
\usepackage{times}
\usepackage{dcolumn}
\usepackage[utf8]{inputenc}
\usepackage[T1]{fontenc}
\usepackage{hyperref}
\hypersetup{
  colorlinks   = true,    % Colours links instead of ugly boxes
  urlcolor     = blue,    % Colour for external hyperlinks
  linkcolor    = blue,    % Colour of internal links
  citecolor    = green      % Colour of citations
}

\usepackage{pdfpages}

\usepackage{changes}

\pdfpageattr {/Group << /S /Transparency /I true /CS /DeviceRGB>>} %necessary for transparent plots

\makeatletter
\AtBeginDocument{\let\LS@rot\@undefined} %needed when including pdf supp_mat at the end
\makeatother

\begin{document}

% 5 proposals
\title{Importance of effective Coulomb interactions for $T_c$ in cuprates}

\author{Jakša Vučičević}
\email{jaksa.vucicevic@ipb.ac.rs}
\affiliation{Scientific Computing Laboratory, Center for the Study of
Complex Systems,\\
Institute of Physics Belgrade, University of Belgrade, Pregrevica 118,
11080 Belgrade, Serbia}

\author{Upendra Kumar}

\affiliation{Scientific Computing Laboratory, Center for the Study of
Complex Systems,\\
Institute of Physics Belgrade, University of Belgrade, Pregrevica 118,
11080 Belgrade, Serbia}
\affiliation{
State Key Laboratory of Critical Mineral Research and Exploration, Institute of Geochemistry, Chinese Academy of Sciences, Guiyang 550081, China
}

\author{Chia-Nan Yeh}
\affiliation{Center for Computational Quantum Physics, Flatiron Institute, New York, NY, 10010, USA}

\author{Miguel A. Morales} 
%\orcidlink{0000-0002-6389-3067}}
\affiliation{Center for Computational Quantum Physics, Flatiron Institute, New York, NY, 10010, USA}

\author{Malte Rösner}
\affiliation{Faculty of Physics, Bielefeld University, 33501 Bielefeld, Germany}

\date{\today}

\begin{abstract}
Cuprate superconductors exhibit the highest observed critical temperatures for superconductivity ($T_c$) at atmospheric pressure.
However, the magnitude of $T_c$ varies significantly between different cuprate compounds.
At present, it is unclear what properties of the crystal structure affect $T_c$ most strongly, yet such an understanding must underpin any efforts toward high-$T_c$ materials design.
To address this issue, we perform a large scale systematic study, employing a combination of data collection, state-of-the-art numerical methods, and statistical analysis.
We start by identifying about 40 different cuprate compounds, and we compile detailed data about their $T_c$'s and crystal structures from literature and the available databases.
Next, using a fully automated procedure, for each compound we compute the density functional theory (DFT) bandstructure and downfold it to two of the most commonly studied low-energy lattice models, namely the single-band Hubbard and the three-band Emery models.
The downfolding is based on the approach of maximally localized Wannier functions (MLWF) and constrained random phase approximation (cRPA).
Finally, we apply a thorough and unbiased statistical analysis to investigate the correlations between the experimentally measured $T_c$'s and the computed parameters of our theoretical models.
Our data indicates that different families of the cuprates follow different trends and that more sophisticated models might be needed to describe all cuprates on the same footing.
Nevertheless, we find that $T_c$ scales well with simple functions of model parameters.
We confirm a previously observed trend that the next-nearest neighbor hopping $t'$ in the single-band model correlates with the experimental $T_c$, and we find that $T_c$ appears to vanish below a finite value of $t'$, in agreement with recent ground-state calculations for the Hubbard model. However, we find that the coupling strength also plays a role: throughout our entire dataset, $T_c$ correlates the most  with the Coulomb coupling on the $p$-orbitals in the 3-band model, highlighting the importance of the oxygen sites in the copper-oxide planes.
\end{abstract}

%\pacs{}

\maketitle

\newcommand{\expv}[1]{\langle #1 \rangle}
\newcommand{\ImG}{\mathrm{Im}G}

\section{Introduction}

The discovery of high-temperature superconductivity has invigorated the efforts to understand the physics of strong electronic correlations\cite{Scalapino2012,Keimer2015,Varma2020}.
The last three decades have seen a significant progress in this respect, in large part based on the theoretical studies of simplified, yet microscopic, low-energy lattice models\cite{Maier2000,Kyung2006,Civelli2008NodalAntinodalTwoGaps,Kancharla2008,Ferrero2009,Gull2013,WuPRB2017,Wu2018,VucicevicPRL2015,VucicevicPRL2021,VucicevicPRB2021,Li2021Nature,Limelette2003,Terletska2011,vucicevic2013,Furukawa2015}.
The simplicity of the models is key, as only the simplest models lend themselves to truly controlled and sometimes numerically-exact theoretical treatments.
Numerical studies of the Hubbard, Emery and $tJ$ 
lattice models appear to converge on the understanding that an emergent antiferromagnetic interaction between neighboring electrons is at the core of the pairing mechanism in the cuprates\cite{Sorella2002,Prelovsek2005,Capone2006,Onufrieva2009,Metlitski2010,Onufrieva2012,Wang2014,Fratino2016,Vucicevic2017,Jiang2021,KowalskiPNAS2021,Dong2022,Roig2022,Chen2025,BacqLabreuil2025}, which is also supported by recent experimental results\cite{OMahony2022}.
However, the question of the magnitude of $T_c$ is a quantitative one that cannot be studied using only generic models; the models need to capture the specifics of the crystal structure that affect $T_c$.
What exactly these specifics might be is one of the central, long-standing questions in the field.

An apparent approach to the study of what determines $T_c$ in the cuprates is to try and make use of the available experimental data and employ regression methods that uncover correlations between $T_c$ and other properties of the material\cite{Feiner1992,Raimondi1996,Pavarini2001,Weber2012,Kim2018,Stanev2018,Hamidieh2018,Lee2021,Wang2023,VucicevicPRB2024}.
However, one never obtains a single result, but rather a multitude of fine-tuned predictor models or phenomenological laws that fit the observations to a varying degree. Identifying among those the ones that symbolize true physical trends or causation is a difficult task.
Here we propose that the most practical way to deal with the ambiguity of statistical approaches is to restrict oneself to studying only those phenomenological laws that can in principle be cross-checked by numerical simulation.
In some previous studies\cite{Feiner1992,Raimondi1996,Pavarini2001,Kim2018,Hamidieh2018}, quantities like spatial distances between atoms and properties of the elements constituting a material have been found to correlate with the $T_c$. Such properties, however, cannot be easily encoded in minimal microscopic physical models,
and their correlations with $T_c$ cannot be easily understood or reproduced by a controlled theory (although some progress in this respect\cite{Vadnais2026arxiv} has been made at the time of writing this manuscript).

In this paper, we aim to investigate the relations between parameters of the simplest microscopic lattice models derived for cuprate compounds and the experimentally measured $T_c$'s in those compounds; any trend that we observe will be possible to cross-check in the future, at least in principle, using, for example, quantum many-body numerical methods. The possibility to cross-check observed experimental trends is essential: if a model truly captures the mechanisms relevant for the magnitude of $T_c$, the experimentally measured $T_c$ should be a uniquely valued function of its parameters, but this alone would not prove the validity of the model; one also needs to be able to reproduce the experimentally observed trends by solving the model and computing $T_c$. The latter is surely difficult, but the knowledge of experimental trends, i.e. how the experimentally measured $T_c$ depends on model parameters, is a prerequisite to even attempt to prove that a model is the correct one for the description of $T_c$ in the cuprates and that the procedure to parametrize the model is correct. Here we have set out to document robust experimental trends so that in future calculations the validity of standard lattice models for the cuprates (as well as the standard approach to their parametrization) can be tested. Similar studies have been carried out in the past\cite{Pavarini2001,Weber2012,Sakakibara2014,Nilsson2019,VucicevicPRB2024,Hirayama2018,Hirayama2019,Moree2022}, but, to the best of our knowledge, they employed much more limited datasets, and they mostly focused on kinetic parameters (with the exception of \cite{Nilsson2019,Moree2022}), whereas here we study the full set of model parameters, including the effective Coulomb coupling constants.

The underlying assumption of our work is that the variation of the (maximal) $T_c$ achieved in different cuprate compounds can be described by 
relatively simple lattice models, and that the full \emph{ab initio} treatment is, ultimately, not needed\cite{LeBlanc2015,Zheng2017}.
Our approach is meaningful if our models can be fully defined using a relatively small number of parameters (say 5-10)
and there is a systematic, well defined procedure to evaluate them for each compound separately.
Formally, lattice models are derived by restricting the degrees of freedom of the system: the entirety of the continuous physical space is replaced by a set of point-like orbitals, and the kinetic energy is encoded in the hopping amplitudes connecting those orbitals, in a manner of a tight-binding model\cite{Mostofi2014}.
The repulsive $\sim 1/r$ Coulomb interaction is replaced by an effective, partially screened interaction, usually short-ranged and formally expressed as a (simplified) Coulomb tensor, with finite diagonal elements\cite{Schuler2013,Aryasetiawan2004}.
However, there is no unambiguous way to derive or parametrize such models.
A common constraint imposed here (as we do in this paper) is that the model should, in the non-interacting limit, reproduce certain low-energy bands of a 
DFT computed bandstructure\cite{Mostofi2014}, at least roughly.
Some of the ambiguity is resolved by invoking the principle of maximally localized Wannier functions\cite{MarzariRMP2012}.
Yet, even with these constraints, the models can still differ in the number and the choice of the orbital degrees of freedom (per unit cell), as well as in the form of the effective interactions.
In particular, parametrization of the effective interactions remains the most problematic step in the downfolding procedure, as it amounts to solving a many-body problem in its own right\cite{Honerkamp2012,Profe2026}.
Here we employ the widely used cRPA approximation\cite{Aryasetiawan2004,Aryasetiawan2006,Springer1998,Miyake2008,Vaugier2012,Shinaoka2015,Scott2024,Chang2024b}, and we consider multiple ways of converting the resulting interactions, which are frequency dependent, into (instantaneous) Hamiltonian terms.
Starting from the undoped, stoichiometric, parent compounds we derive the two most standard models, namely the single-orbital Hubbard and the three-orbital Emery models, and we restrict ourselves to density-density interactions acting only within a single unit cell. 
We understand doping as a shift in the chemical potential, which we do not consider in the subsequent analysis.

Our results reveal correlations between $T_c$ and the parameters of our models, across the entire dataset. However, the experimental $T_c$ cannot be fully described as a function of 1-2 model parameters.
Better scaling is observed when considering different categories of compounds separately, each defined by distinct features of the crystal or the band structure.
Overall, in the single-band model, the strongest correlations are observed with the next nearest-neighbor hopping-amplitude $t'$, its ratio to the nearest-neighbor hopping $t'/t$, as well as functions of the strength of the on-site interaction $U$ and the effective next-nearest neighbor antiferromagnetic interaction $J'\equiv t'^2/U$. These statistically observed trends are in line with previous works that employed similar regression methods\cite{Raimondi1996,Pavarini2001,Weber2012,Nilsson2019}, but also those computing the ground state of the square lattice Hubbard model\cite{Harland2016,JiangPRB2024,Qin2020,Xu2024}: $T_c$ is expected to vanish below a certain value of the ratio $|t'/t|\sim 0.1$ and this is precisely what we see.
In the 3-band model, we find that $T_c$ correlates the most with the parameters of the local interaction on the oxygen sites.
This supports the idea that oxygen sites play an important role\cite{KowalskiPNAS2021,Rybicki2016,Hansmann2014} and suggests that the common approach of neglecting interactions on the $p$-orbitals\cite{Kent2008,Weber2008,deMedici2009,Weber2012,KowalskiPNAS2021,Malcolms2024arxiv, VucicevicPRB2024,Tseng2025,StCyr2025,Vucicevic2026arxiv} might not be fully justified.
In fact, the best single parameter to describe $T_c$ across our entire dataset is the coupling constant on the oxygen site of an effective Emery-Holstein model, where local cRPA frequency-dependent interactions are approximated by couplings to dispersionless bosonic modes. Our results thus suggest that the effects of retardations in the Coulomb interactions are important and that they can be modeled using Holstein-like\cite{Mahan2000book} terms in an augmented Hamiltonian.

The rest of the paper is organized as follows. The sections follow the order of steps we have performed in our work.
Section~\ref{sec:data_collection} explains the starting point of our study, which is a collection of the experimental $T_c$ and the crystal structure data for various cuprates.
Our final data collection is fully documented in Appendix~\ref{app:Tc_table}.
In Section~\ref{sec:ab_initio} (and Appendix~\ref{app:dft_and_wannierization}) we present details of our DFT calculations.
We discuss the properties of crystal structures and their relation to DFT band-structure and $T_c$ in Section~\ref{sec:crystal_and_bandstructure_analysis}.
In Section~\ref{sec:downfolding}, we describe the details of our downfolding procedure. The downfolding step is further detailed in Appendix~\ref{app:dft_and_wannierization}, and the cRPA step is discussed in Appendix~\ref{app:cRPA}, where it is also compared with previous works. The results of cRPA are discussed in Section~\ref{sec:crpa_analysis}, while full model parametrization is illustrated in Appendix~\ref{app:orientation}.
The correlations of model parameters with experimentally measured $T_c$ are investigated in Section~\ref{sec:correlations_with_Tc} with further details given in Appendix~\ref{app:parameters}.
In Section~\ref{sec:conclusion} we summarize our conclusions, discuss the limitations of our study and envision prospects for future work.
Full parametrization of our models and the full results of our statistical analyses are available as Supplemental Material\cite{SMrepo}.

\section{Data collection}
\label{sec:data_collection}
We identify about 40 stoichiometric parent compounds for which $T_c$ has been reported (with or without doping) in at least one peer-reviewed publication.
In many cases, different doping schemes are possible, but they sometimes lead to similar maximal $T_c$ (e.g. in the case of La$_2$CuO$_4$, doping by Sr and O lead to similar maximal $T_c$\cite{trofimov1994growth,tarascon1987superconductivity,chaillout1989crystal}, but doping with Ba leads to slightly lower $T_c$\cite{Guguchia2023}; doping with Ce is substantially different, because it represents electron-doping, which usually leads to lower $T_c$'s\cite{naito2000superconducting}).
For many of the compounds we are also able to find experimental results for the space group of the crystal structure; in any case, we find the corresponding crystal structures in The Materials Project database\cite{Horton2025,Jain2013} and a few in Crystallography Open Database\cite{Vaitkus2023}. The $T_c$, space-group data and the crystal structure ID's are tabulated in Appendix~\ref{app:Tc_table} (Table~\ref{tab:Tc}), where we also illustrate the conventional unit cells for all crystal structures (Fig.~\ref{fig:collage}).

It is worth noting that some compounds go through structural phase transitions - the details of the crystal structure can change depending on temperature or doping.
As we are interested in treating a large amount of compounds on the same footing, such detailed behaviors are outside of the scope of our study - rather, if multiple crystal structures are relevant for a given compound, we focus on the ones with the smallest primitive unit cell.

\section{Ab initio calculation}
\label{sec:ab_initio}
The next step is to obtain the DFT bandstructure for each compound. To do this we make use of \texttt{Quantum Espresso}\cite{giannozzi2009,Giannozzi2017}.
All calculations are performed with the same numerical parameters, as explained in detail in Appendix~\ref{app:dft_and_wannierization}.
We repeat calculations with two different sets of pseudopotentials [SSSP\cite{prandini2018precision} and ONCV\cite{Hamann2013}(with DOJO\cite{vanSetten2018} for $f$-elements)], and confirm that the results are similar with both.
We compute orbital-projected bandstructures, as well as the corresponding partial densities of states to inform the systematic derivation of Wannier functions at a later stage.

\begin{figure*}[ht!]
\centering
\includegraphics[width=1.0\textwidth, trim=3.0cm 0 3.0cm 0, clip]{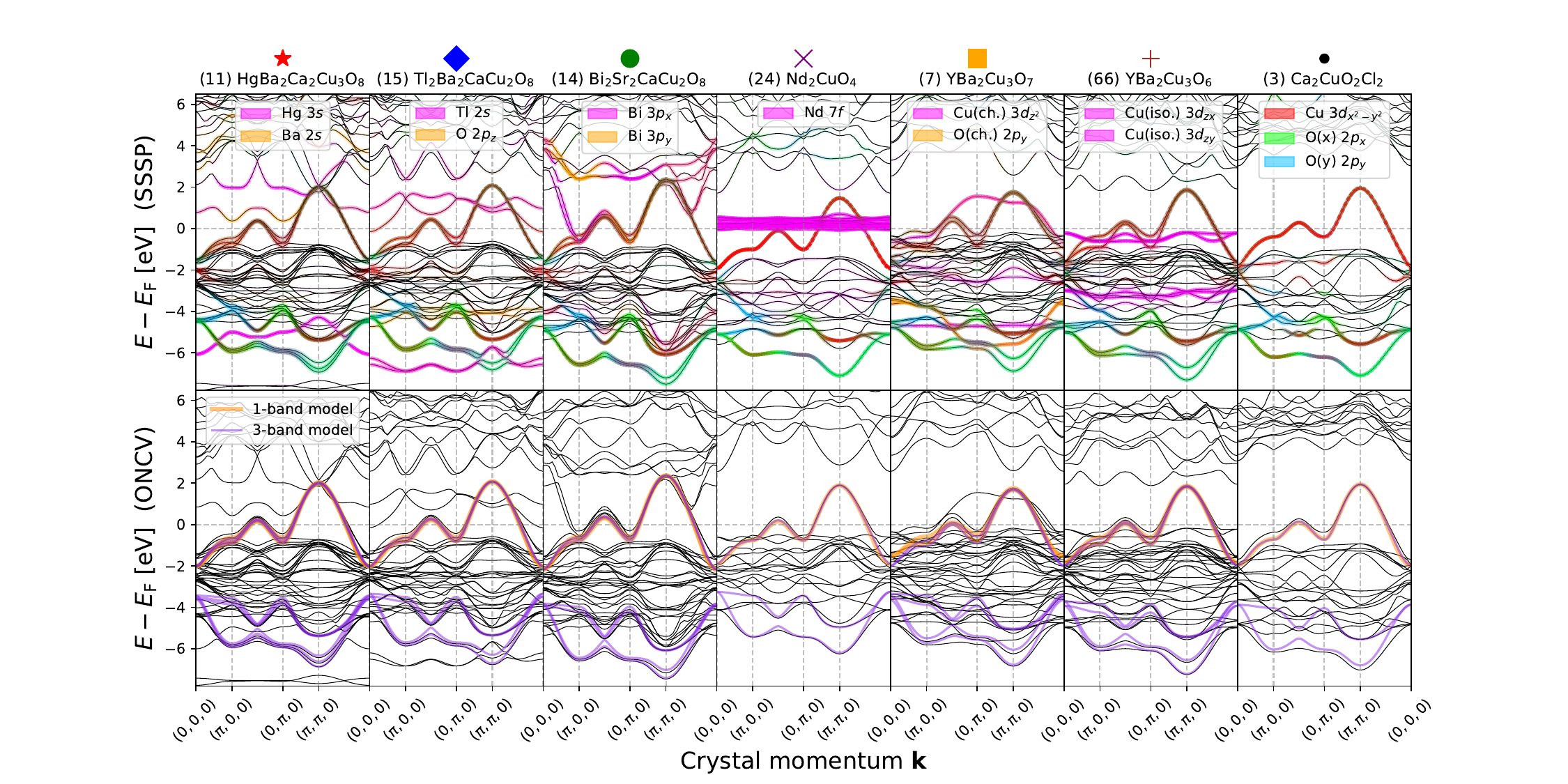}
\caption{Examples of our DFT and downfolded bandstructures. Crystals are oriented so that copper-oxide layers are parallel to the $xy$-plane, and the projections of the Cu-O bonds on this plane point in the $x$ and $y$ directions. Bandstructures are shown in the $k_z=0$ plane, and we have checked there is little variation with $k_z$. Top row: DFT bandstructures computed with SSSP\cite{prandini2018precision} pseudopotentials, with color-coded orbital character; red, skyblue and lime correspond to Cu-$3d_{x^2-y^2}$, O-$2p_x$, and O-$2p_y$ orbitals, of the coppers and oxygens forming the copper-oxide layers; magenta and orange correspond to orbitals outside of the copper-oxide layers found to contribute to bands near the Fermi level. Bottom row: DFT bandstructures computed with ONCV+DOJO\cite{Hamann2013,vanSetten2018} pseudopotentials, with overlayed downfolded bandstructures for the single-band and three-band models. Each column corresponds to a different category of compounds which we define by distinct features of the crystal and bandstructure; here we illustrate each category with a single example, but other compounds from the same category have a similar DFT bandstructure, in ways discussed in the main text. The symbols above the panels will be used to denote each category in the rest of the plots. For reference, the maximal $T_c$'s of the compounds shown are (left to right): 135K, 110K, 98K, 32.5K, 93K, 90K and 26K. 
}
\label{fig:dft_categories}
\end{figure*}

\section{Crystal- and band-structure data analysis}
\label{sec:crystal_and_bandstructure_analysis}

All crystal structures that we consider contain at least one copper-oxide CuO$_2$ layer per primitive unit cell, but some have as many as four such layers.
When there are two layers, they are crystallographically equivalent; when there are 3 or 4 layers, then the inner and the outer layers are distinct; in our dataset, the maximum number of crystallographically distinct layers is therefore 2.
In a given CuO$_2$ layer, the copper atoms form a square (or a slightly rectangular) lattice, and oxygens are placed in the middle of each pair of neighboring coppers. In some cases, oxygens are slightly out of the plane formed by the coppers. For the sake of simplicity, we only consider crystal structures with no enlargement of the unit cell, i.e. we always have a single copper atom per CuO$_2$ layer per (primitive) unit cell.
In a number of crystal structures, copper oxide CuO chains running parallel to the CuO$_2$ planes are present. In one compound, there is a pure copper layer with no oxygen atoms in it.

By visually inspecting the obtained band-structures, we observe that they are similar in some respects (representative examples are shown in Fig.~\ref{fig:dft_categories} and the full set of results is available in the Supplemental Material\cite{SMrepo}).
As expected, at the Fermi level one always finds a composite band whose orbital character is dominated by Cu 3$d$ and O 2$p$ states from the atoms forming the CuO$_2$ layers. Specifically, the band crossing the Fermi level has mainly Cu 3$d_{x^2-y^2}$ character and, in certain regions of the Brillouin zone (BZ), also a significant contribution from O 2$p_x$ and 2$p_y$ orbitals pointing toward the nearest Cu atoms. The main contribution of these oxygen orbitals appears in the two bands at the bottom of the composite band, where some Cu 3$d_{x^2-y^2}$ character is also present \cite{Mattheiss1987,Emery1987,Varma1987,Hansmann2014}.
In compounds with multiple copper-oxide layers per primitive unit cell ($N_\mathrm{layers}>1$), the bands come in groups of $N_\mathrm{layers}$ nearly degenerate bands. The distinct copper-oxide planes do not differ by much and only weakly hybridize with each other.

The bandstructures for different compounds, however, differ in some important details. In some cases, there will be additional bands crossing or coming close to the Fermi level. We categorize the compounds by the shape and the presence of these additional bands, and find that each category corresponds to a particular feature of the chemical composition or the crystal structure. We define 7 categories, and illustrate them each with one example in the top row of Fig.~\ref{fig:dft_categories}.
The first 6 categories (counting left-to-right) all display the additional bands, but their orbital character comes from different orbitals. The orbital character is color-coded on top of the bands; all plots show the Cu 3$d_{x^2-y^2}$ and O 2$p_{x/y}$ character in the red, skyblue and lime color; the orbital character relevant for the additional bands is colored orange and magenta. In the first four groups, the additional bands come from the orbitals of atoms in the buffer layers, found in between the copper-oxide layers. The shape of the additional bands is indicative of the presence of certain elements, which defines the different families of the cuprates. In the fourth group, the additional bands are dispersionless, and they correspond to $f$-orbitals, which are highly localized.
Here we emphasize that the DFT bandstructure cannot capture all the true spectral features of the cuprates; especially in the case when $f$-elements are present, one expects that, in reality, the dispersionless $f$-bands will be split by strong effective interactions and that they will not be present anywhere near the Fermi level.
For that reason, in DFT calculations, the $f$-orbitals are often kept in the frozen core, i.e. they are absorbed by the pseudopotentials. We show the result of such a DFT calculation on the plots in the bottom row (using DOJO\cite{vanSetten2018} pseudopotentials for $f$-elements), and show that the band structure is then much more similar to the 7th group, where no additional bands are present near the Fermi level.
In the 5th group, the additional band comes from the CuO chains present in those crystal structures, in particular the Cu 3$d_{z^2}$ and the O 2$p_{x/y}$ orbitals pointing along the chain. The 6th group comprises a single crystal structure where we observe a layer of pure copper, as already mentioned; the additional bands just below the Fermi level come precisely from the Cu $3d_{zx/zy}$ orbitals of those coppers and are weakly dispersive.

\begin{figure*}[ht!]
\centering
\includegraphics[width=1.0\textwidth, trim=0.0cm 0 0cm 0, clip]{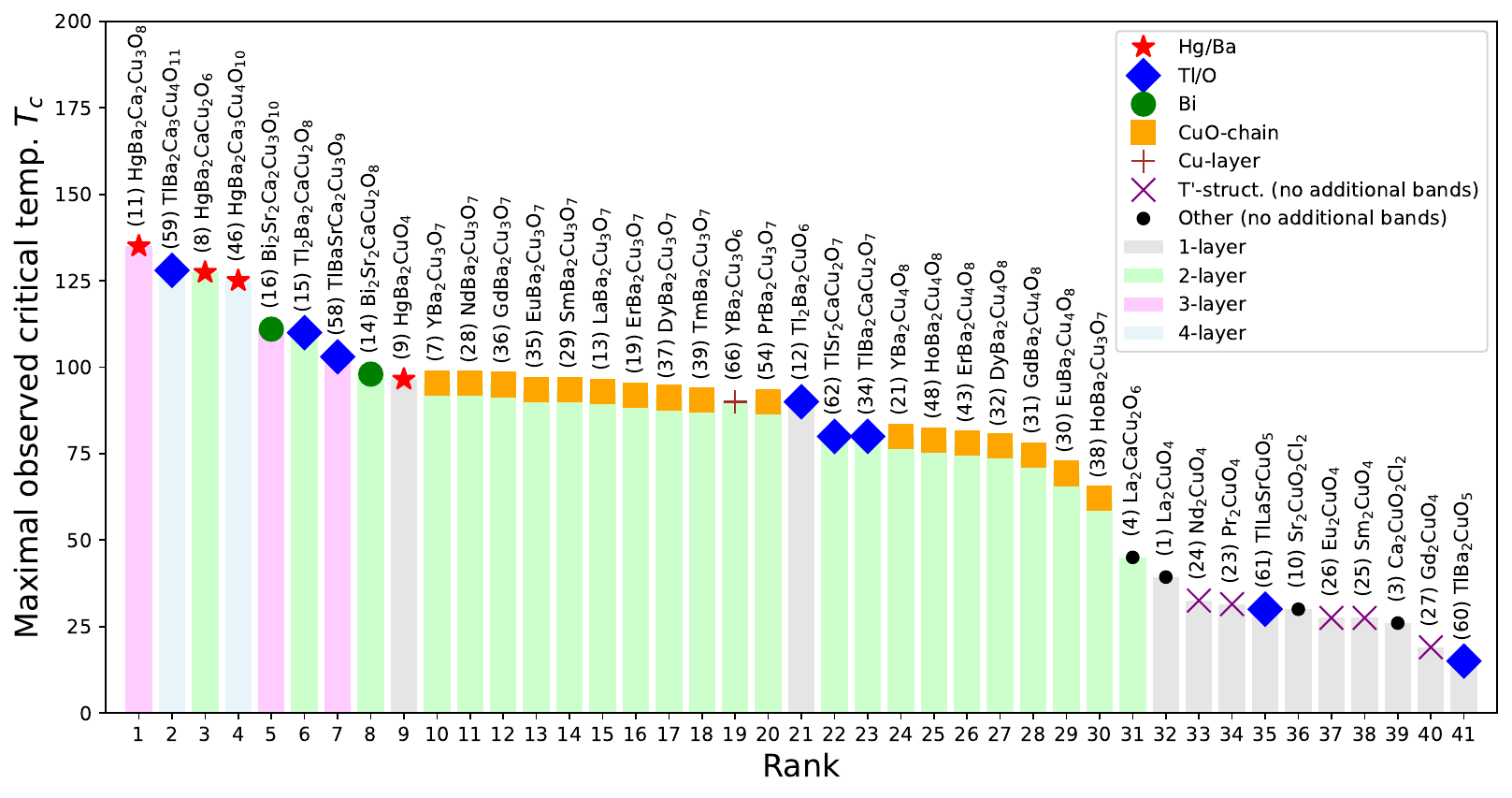}
\caption{
Maximal observed $T_c$ (regardless of doping)
in our dataset of parent compounds.
Symbol denotes the category of the compound (based on the presence and the orbital character of additional bands near the Fermi level in the DFT band structure; the categories are illustrated in different columns of Fig.~\ref{fig:dft_categories}
; color of the bar denotes the number of CuO$_2$ layers per (primitive) unit cell.
}
\label{fig:Tc_vs_category}
\end{figure*}

To investigate whether these broad characteristics of the DFT bandstructure can already be related to the magnitude of $T_c$ we plot in Fig.~\ref{fig:Tc_vs_category} the highest observed $T_c$ in each of our compounds; we order the compounds by the $T_c$ and we denote them by different symbols, depending on the category they belong to.
We observe that the compounds with no additional bands in the DFT bandstructure (including the $f$-element group) have by far the lowest $T_c$. The compounds with CuO chains display remarkably little variation of $T_c$, and cluster around 80 K. The compounds with $T_c>50$ K all feature additional bands close to the Fermi level in the DFT band structure. In our dataset, the biggest spread of $T_c$ is observed in the thallium based cuprates. A similarly good perspective can also be obtained by considering $N_\mathrm{layers}$ for each of the compounds. As has been discussed in literature\cite{BacqLabreuil2025}, 3-layer compounds display the highest $T_c$'s. However, we observe that there are single-layer compounds approaching $T_c\sim100$ K, and the compound with the third highest $T_c$ in our dataset has $N_\mathrm{layers}=2$; in the Tl/O category, the two highest $T_c$'s are found in 4- and 2-layer compounds; there are clearly features other than the number of CuO$_2$ layers per primitive unit cell that contribute to $T_c$. Interestingly, the effects of multilayeredness might be intertwined with the existence of additional bands coming from outside of the copper-oxide planes, which was discussed in the recent Ref.~\onlinecite{BacqLabreuil2025} (see also Section~\ref{sec:interlayer}).

\section{Downfolding}
\label{sec:downfolding}

The next step is to evaluate the parameters of the effective models for the compounds in our dataset. 
We use the DFT results (computed using ONCV+DOJO pseudopotentials) to 
derive for each compound a single-band (Hubbard) and a three-band (Emery) model. The downfolding onto the two models is performed in two steps:
first we perform a wannierization, which gives us direct access to the tight-binding matrix, and then we do a constrained random phase approximation (cRPA) calculation, which yields effective Coulomb coupling amplitudes, based on the given wannierization result.

We repeat the wannierization step using two separate implementations (\texttt{Wannier90}\cite{Mostofi2014} and \texttt{RESPACK}\cite{Nakamura2016,Nakamura2021}) of the Souza-Marzari-Vanderbilt algorithm\cite{Souza2001} [the maximally localized Wannier functions (MLWF) approach for the entangled bands]
and confirm that the results obtained are the same with both (in some cases, one of the implementations failed to converge the MLWFs, while the other was successful; we have found no cases where the two implementations converged to different results).

The main input parameters for the wannierization are the inner and outer energy windows, and the initial guess for the Wannier orbitals.
For the Hubbard model, the initial guess are the atomic Cu 3$d_{x^2-y^2}$ orbitals of the coppers in the copper-oxide layers; for the Emery model we also include the appropriate O $2p_{x/y}$ orbitals. The total number of Wannier orbitals (orbitals per unit cell in the lattice model) is therefore $N_\mathrm{layers}$ for the Hubbard model and $3N_\mathrm{layers}$ for the Emery model.

We did not use the inner (frozen) energy windows even though in some cases it can improve the agreement of the downfolded bandstructure with the DFT bandstructure.
The reason is that the frozen energy window cannot be applied in a consistent and unbiased manner across the entire dataset. 
As already mentioned, the DFT bandstructure for a large number of compounds features additional bands coming close to, or even crossing the Fermi level.
When the additional bands overlap with those bands that we wish to wannierize, this poses a problem: it is a hard constraint of the wannierization algorithm that the number of bands appearing in the frozen window (at any $k$-point) is not larger than the number of (initial-guess) wannier orbitals. Therefore, for a large number of compounds, one cannot use the frozen window in a physically meaningful way. When there are additional bands crossing the Fermi level, using even a tiny frozen window around $E_\mathrm{F}$ leads to a nonphysical wannierized bandstructure that describes different DFT bands in different parts of the Brillouin zone, "switching" abruptly from one to the other as momentum space is traversed. For a smaller number of compounds where the additional bands are not crossing the Fermi level, a narrow frozen window can be meaningfully set in the range of energies (around $E_F$) where only the band we wish to wannierize appears.
However, the results would depend on the precise choice of the frozen window, while its maximal possible width varies greatly between the compounds (as set by the position of bands closest to the Fermi level). Overall, we conclude that using the frozen window is only possible for a smaller number of compounds, and it would necessarily introduce an undesirable degree of variability in the treatment of the compounds, which would be potentially problematic for the subsequent statistical analyses that we wish to perform.

The outer energy windows we set automatically, based on the energy-spread of the partial density of states, as obtained from projection to the initial orbitals. 
This avoids any human intervention and thus generates bias-free sets of consistently constructed Wannier orbitals.
Those are given by the Wannier functions $\psi_i^\alpha(\mathbf{r})$ centered at site $i$ and with orbital character $\alpha$.
The hopping matrix elements are then obtained according to $t^{\alpha \beta}_{ij} = \langle \psi_i^\alpha | \hat{H}_{DFT} | \psi_j^\beta \rangle$, which also allows us to calculate the downfolded bandstructures for both models. The latter are illustrated in Fig.~\ref{fig:dft_categories} and the full results are available in the Supplemental Material\cite{SMrepo}.

Following the Wannierization, the cRPA calculations are performed using the CoQuí software library~\cite{CoQui_GitHub,THCGW_Yeh2024}; see implementation details in Appendix~\ref{app:cRPA}. 
The cRPA algorithm can be understood as an approximation of a more general, rigorous formalism\cite{Profe2026} in which a subset of the single-particle degrees of freedom (d.o.f.) are integrated out;
in our case, the goal is to integrate out all d.o.f. but the (maximally-localized) Wannier orbitals. 
In general, integrating out d.o.f. involved in two-body interactions (as is presently the case) leads to an effective-action description of the system that features all possible time-dependent many-body terms. The conventional wisdom that underlies the lattice-model approach is that the main correlation effects come from short-range two-body terms in the effective action, and that the rest can be neglected. 
At the level of cRPA approximation, the effective interactions are obtained as bosonic-Matsubara-frequency dependent and tensor-valued amplitudes ${\cal U}_{\alpha\beta\gamma\delta}(i\nu)$ (the indices $\alpha\beta\gamma\delta$ enumerate orbitals within a single primitive unit cell of the effective model).
However, it is less obvious whether or how one can transform retarded interactions into instantaneous, Hamiltonian terms.
The most common approach is to take the static component ${\cal U}_{\alpha\beta\gamma\delta}(i\nu=0)$ as the instantaneous-coupling constant, but other options are possible\cite{Casula2012,Scott2024,PauliPRB2025}. It is not even necessary to approximate the interactions as instantaneous: many-body treatments of lattice models with retarded interactions are possible; however, keeping 3-body or higher-order interactions in the model would preclude solutions using most implementations of the standard numerical approaches. In our work, we focus on the density-density terms ($\alpha=\beta$, $\gamma=\delta$, ${\cal U}_{\alpha\beta}(i\nu)\equiv{\cal U}_{\alpha\alpha\beta\beta}(i\nu)$) and consider the entire frequency dependence of the effective interactions. We again emphasize that both the wannierization and cRPA steps are performed for all CuO$_2$ layers simultaneously - the indices $\alpha$ and $\beta$ go over all the relevant orbitals in all the CuO$_2$ layers.  

An important technical subtlety in the implementation of cRPA approach concerns the way one differentiates between the screening channels within the "target" low-energy subspace and those from the "rest" space\cite{Miyake2009,Honerkamp2018,Kaltak2025_cRPA_spectral}. 
When the target atomic characters of the MLWFs are strongly entangled with multiple DFT bands, the Wannierization does not necessarily reproduce the original DFT bands perfectly. 
As a result, the Bloch states that diagonalize the Kohn-Sham Hamiltonian may differ from the eigenstates of the downfolded Wannier Hamiltonian.
If the constrained subspace is defined directly from the latter, this mismatch can leave residual low-energy polarization channels in the cRPA calculation, leading to spurious screening and a significant underestimation of the screened interaction~\cite{Miyake2009,Kaltak2025_cRPA_spectral}. 
To avoid this artifact, we instead exclude, at each $\mathbf{k}$ point, the Kohn-Sham eigenstates with the largest overlap with the MLWF subspace, rather than exclude the subspace spanned by the MLWFs themselves. This strategy is consistent with the modified cRPA algorithm recently proposed by \emph{Kaltak et al.}~\cite{Kaltak2025_cRPA_spectral}.

Our cRPA results are illustrated in Appendix~\ref{app:cRPA}, and the full results are available in the Supplemental Material\cite{SMrepo}.

We emphasize that we were unable to complete all the procedures for all the compounds; in some cases and for various reasons, we could either not complete the DFT, the wannierization part or the cRPA part, depending also on the model. This means that the sets of compounds for which we do the analysis will be slightly different for the single-band and the three-band models, and slightly smaller than the original 41-compounds dataset.

\section{Analysis of the cRPA results}
\label{sec:crpa_analysis}

The cRPA results for all compounds are qualitatively similar and as expected. The amplitude ${\cal U}_{\alpha\beta}(i\nu)$ is the smallest at $i\nu=0$, and it grows slowly with increasing frequency; it finally saturates at the bare Coulomb value $V_{\alpha\beta} \sim \int \mathrm{d}\mathbf{r}\mathrm{d}\mathbf{r'} \psi_\alpha^*(\mathbf{r})\psi_\alpha(\mathbf{r}) \frac{1}{|\mathbf{r}-\mathbf{r}'|}\psi_\beta^*(\mathbf{r}')\psi_\beta(\mathbf{r}')$ which can be evaluated directly from the wannierization results, i.e. from the knowledge of the Wannier functions $\psi_\alpha(\mathbf{r})$.
Quantitatively, however, we observe that there is a lot of variation in the effective interactions across our dataset.
Some examples of results are given in Appendix~\ref{app:cRPA}, while full results are given in the Supplemental Material\cite{SMrepo}.
In the Hubbard model, we have only one orbital per unit cell, and we only consider the local density-density interaction; at $\nu=0$ we get values roughly in the range 0.5-3.5eV, at $\nu\sim 5$eV, the values are around 4-5.5eV, and the bare-Coulomb $V$ mostly is in the range 14-16eV. We can compare these values to the onsite density-density interactions on the Cu $d$-like orbital in the Emery model, and see that those are larger: at $\nu=0$ values are in the range 3-6.3eV, at $\nu\sim 5$eV, roughly 7.5-9.5eV, and bare Coulomb is around 25.5eV.
In fact, it is expected that the interactions in the single-band model will be smaller than in the three-band model: reproducing the DFT bandstructure at the Fermi level with fewer orbitals requires that the corresponding Wannier orbitals are more delocalized, thus the bare interactions will be smaller. This also translates to the partially screened cRPA Coulomb matrix elements.
Analogously, in the three-orbital model, the bare Coulomb interactions $V$ on the $p$-like orbitals are universally smaller than that on the $d$-like orbitals, by about 30\% (see Appendix~\ref{app:dft_and_wannierization}). However, at low frequencies, the difference between the on-site interactions on the $d$ and $p$-like orbitals is much smaller, and in a number of cases we find that the interactions are stronger on the $p$-orbitals, indicating a relatively stronger cRPA screening to the $d$-orbital Coulomb matrix elements. This is somewhat surprising and at odds with some previous estimates\cite{Hanke2010,Hansmann2014}, and immediately suggests that the effective interactions on the oxygen $p$-orbitals might be more important than previously thought; as already mentioned, in many Emery-model calculations, the interactions on the $p$-orbitals are entirely neglected\cite{Kent2008,Weber2008,deMedici2009,Weber2012,KowalskiPNAS2021,Malcolms2024arxiv, Tseng2025,StCyr2025}.

\section{Correlations with $T_c$}
\label{sec:correlations_with_Tc}

We arrive at the central part of this work. We have generated a dataset of parameters of the effective Hubbard and Emery models for a large number of cuprate compounds; now we will perform regression analysis to try and uncover relations between the experimentally measured $T_c$ and the model parameters we have computed for those compounds. The analysis will be performed independently, but fully analogously, for the two models.

\subsection{Parametrization of the models}
\label{sec:parametrization}

\begin{figure}[ht!]
\centering
\includegraphics[width=1.0\columnwidth, trim=0.0cm 0 0cm 0, clip]{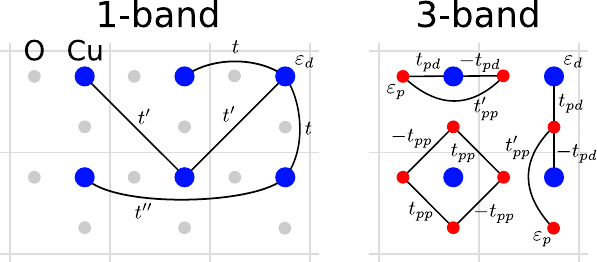}
\caption{
Illustration of the hopping amplitudes in the copper-oxide layers considered in this work; left: single-band model; right: three-band model.
}
\label{fig:tb_illustration}
\end{figure}

Our models can be described with the following general action:
\begin{eqnarray}\label{eq:imp_action}
 S &=& \sum_{ij,\alpha\beta,\sigma}\int \mathrm{d}\tau \bar{c}_{\alpha,i,\sigma}(\tau) [\partial_\tau -\mu + t^{\alpha,\beta}_{ij}] c_{\beta,j,\sigma}(\tau) \\ \nonumber
 &&+
 \sum_{i,\alpha\beta,\sigma\sigma'}\iint \mathrm{d}\tau\mathrm{d}\tau' \frac{{\cal U}_{\alpha\beta}(\tau-\tau')}{2}
 n_{\alpha,i,\sigma}(\tau)
 n_{\beta,i,\sigma'}(\tau')
\end{eqnarray}
where $\bar{c}/c$ are Grassmann fields, $n=\bar{c}c$, $i$ denotes unit cells, $\alpha,\beta$ denote orbitals within a unit cell (of which we will have one or three per CuO$_2$ layer), $\sigma$ denotes spin projection, and $\tau$ is imaginary time. The $t_{ij}^{\alpha\beta}$ is the tight-binding matrix; the diagonal elements correspond to on-site energies, $\varepsilon_\alpha \equiv t_{ii}^{\alpha\alpha}$; in the single-band model we further define $t$, $t'$ and $t''$ as the nearest, next-nearest and third-neighbor hopping, respectively. In the 3-band model, we define $t_{pd}$ as the hopping between nearest-neighbor copper and oxygen orbitals. We also define $t_{pp}$ and $t'_{pp}$ as hoppings between the nearest and next-nearest oxygen orbitals, respectively. This is illustrated in Fig.~\ref{fig:tb_illustration}.
The effective Coulomb coupling in our models is retarded, and set by an imaginary-time dependent coupling amplitude ${\cal U}_{\alpha\beta}(\tau)$. The coupling also contains a purely instantaneous component, equal to the bare Coulomb value. We can therefore write
\begin{equation}
{\cal U}_{\alpha\beta}(\tau)=V_{\alpha\beta}\delta(\tau)+\tilde{\cal U}_{\alpha\beta}(\tau).
\end{equation}
In Matsubara-frequency domain the instantaneous component amounts to an overall real-valued shift, ${\cal U}_{\alpha\beta}(i\nu)=\mathrm{FT}[\tilde{\cal U}_{\alpha\beta}(\tau)]+V_{\alpha\beta}$.
In the case we only choose to keep instantaneous interactions, we can formulate the model using the Hamiltonian
\begin{eqnarray}
H &=&  \sum_{ij,\alpha\beta,\sigma}c^\dagger_{\alpha,i,\sigma}[-\mu + t^{\alpha,\beta}_{ij}] c_{\beta,j,\sigma} \\ \nonumber
 &&+
 \sum_{i,\alpha\beta,\sigma\sigma'} \frac{U_{\alpha\beta}}{2}
 n_{\alpha,i,\sigma}
 n_{\beta,i,\sigma'}
\end{eqnarray}
where $c^\dagger/c$ are creation/annihilation operators.
There is no clear prescription how to estimate the instantaneous coupling constant $U_{\alpha\beta}$ based on the ${\cal U}_{\alpha\beta}(\tau)$ that we obtain from cRPA\cite{Casula2012,Scott2024,PauliPRB2025}. One possibility is to consider the zero-frequency value, $U={\cal U}(i\nu=0)$. We also consider the possibility that a meaningful choice is ${\cal U}(i\nu)$ at some finite Matsubara frequency, as it corresponds to an integral of the imaginary part over a range of real frequencies, which would then capture the fact that the transfer frequency in e-e collisions takes on a range of values, roughly set by the full bandwidth. We also consider the possibility that $U$ equals the bare Coulomb value $V$, but one generally expects that effective interactions in lattice models will be less than that, due to screening.

The full list of parameters that we consider for each model is presented in the Appendix~\ref{app:parameters}.
We supplement the basic parameters for each model by some simple, but potentially relevant functions of certain pairs of those parameters.
Namely, in the Emery model, we are primarily interested in the difference between the onsite energies on the $d$ and $p$ orbitals, $\varepsilon_d-\varepsilon_p$, and we will not consider $\varepsilon_d$ and $\varepsilon_p$ separately.
The energy difference $\varepsilon_d-\varepsilon_p$ is one possible definition of the charge transfer gap (CTG)\cite{Zaanen1985,Weber2012}; CTG has been often considered as very relevant for $T_c$, since in the Emery model it controls the effective antiferromagnetic (AFM) interaction\cite{KowalskiPNAS2021,Vadnais2026arxiv}.
In the single-band model we will also be considering the following ratios: $J=t^2/U$, the effective nearest-neighbor (n.n.) AFM interaction (and similarly the next-n.n. $J'=t'^2/U$); $U/t$, the strength of the interaction in the units of the n.n. hopping (and similarly for various interactions and hopping amplitudes in the 3-band model); $t'/t$, the ratio between the n.n. and the next-n.n. hopping; $J'/J=t'^2/t^2$, the ratio between effective n.n. and next-n.n. AFM interactions. For all the effective interactions, we will consider separately the amplitudes at Matsubara frequencies $\nu=0$ and $\nu\approx 5$eV, as well as the bare-Coulomb value. For the sake of readability, the Matsubara frequency will be indicated in parenthesis without physical units (and also without the imaginary unit), and the parameters derived from bare Coulomb values will be denoted without any parentheses, although this strictly corresponds to taking the Matsubara frequency to infinity [e.g. ${\cal U}(5)\equiv{\cal U}(i5\mathrm{eV})$, $J\equiv t^2/V$, $J(0)\equiv t^2/{\cal U}(0)$, $V={\cal U}(i\nu\rightarrow\infty)$ etc.].

We emphasize that none of the tight-binding parameters have been corrected for the double counting of interaction effects. 
The estimation of double counting corrections\cite{Honerkamp2018,Haule2015} is an outstanding problem in the formulation of lattice models, when this is done by downfolding a DFT bandstructure (as we do here). Some of the effects of the interactions are already included at the level of DFT, in the form of an instantaneous (frequency-independent) self-energy that is absorbed by the bandstructure. In general, to avoid the double counting of interaction effects in the lattice-model solution, one has to correct the tight-binding parameters obtained from wannierization. However, the correction is not well known and will in general depend on the choice of interactions to be kept in the model, as well as their coupling constants; it might also depend on the choice of the (approximate) method which is to be used to solve the model. If only the local density-density interactions are kept in the model, then only the on-site energies need to be corrected. In the case of the Hubbard model, the double counting is therefore not an issue, as the correction can be absorbed in the chemical potential; 
in the 3-band Emery model, however, $\varepsilon_d-\varepsilon_p$ would need to be corrected. If also the non-local $U_{dp}$ terms are kept, then also $t_{pd}$ would need to be corrected. As already stated, the correction depends on the specific version of the model and the values of instantaneous coupling constants, and those we do not yet know (they can be taken from cRPA results in multiple ways, as discussed above). Therefore, using the corrected parameters in our statistical analysis would entail doing it separately for each version of the model (different choices of interaction terms and different choices of instantaneous coupling constants), but this is beyond the scope of the present work.
Nevertheless, in the Supplemental Material\cite{SMrepo}, we document for the Emery model all the intra-cell density-density interactions, and we document the DFT single-particle density-matrix $\langle c^\dagger_\alpha c_\beta \rangle$ that can be used to construct double-counting corrections for the tight-binding parameters in a simple approximation\cite{Sheng2022}, for whatever version of this model is constructed based on our wannierization+cRPA data.

As already mentioned, in some compounds the CuO$_2$ layers do not have the full symmetry of the square lattice. In those cases hopping amplitudes in the $x$ and $y$ directions will not be the same; we will then consider values averaged over the $x$ and $y$ directions. In the Emery model, we will similarly average the non-local ${\cal U}_{dp}$ interactions, as well as the local interactions on the $p_x$ and $p_y$ orbitals.

In the cases when $N_\mathrm{layers}>2$, we consider the results for the inner and outer layers as separate data points, as was done in some previous studies\cite{Weber2012, VucicevicPRB2024}. 
Our primary goal is to study parameters for the minimal, single-layer models of the cuprates that are usually solved using numerical many-body methods. Multilayer models are more difficult to solve, unless additional approximations are performed to decouple the layers (e.g. in cluster dynamical mean-field theory one might solve 2D clusters and neglect interlayer correlations\cite{Harland2019, BacqLabreuil2025}).
Nevertheless, it is interesting to consider also the interplane couplings - both the hopping amplitudes and the density-density interactions between the adjacent CuO$_2$ planes.
The interplane hopping amplitudes can be extracted from our wannierization (it being done for all layers simultaneously), and we will analyze them separately in Section~\ref{sec:interlayer}. In principle, the interplane density-density interactions can also be computed within cRPA.
The interplane couplings, however, are smaller and must be more comparable to the uncertainties of the wannierization and cRPA; thus, any observed trends related to inter-layer parameters are more difficult to trust. We follow the same line of reasoning in the case of longer-range (intraplane) hopping amplitudes and interactions, and avoid investigating their correlations with the $T_c$. Although a very recent numerical study of the Emery model\cite{Jacob2026arxiv} suggests that longer range hoppings might affect $T_c$, their amplitudes are small and difficult to extract from DFT with certainty, and they are unlikely to make the dominant contribution to the $T_c$. 

\begin{figure*}[ht!]
\centering
\includegraphics[width=1.0\textwidth, trim=0.0cm 0 0cm 0, clip, page=1]{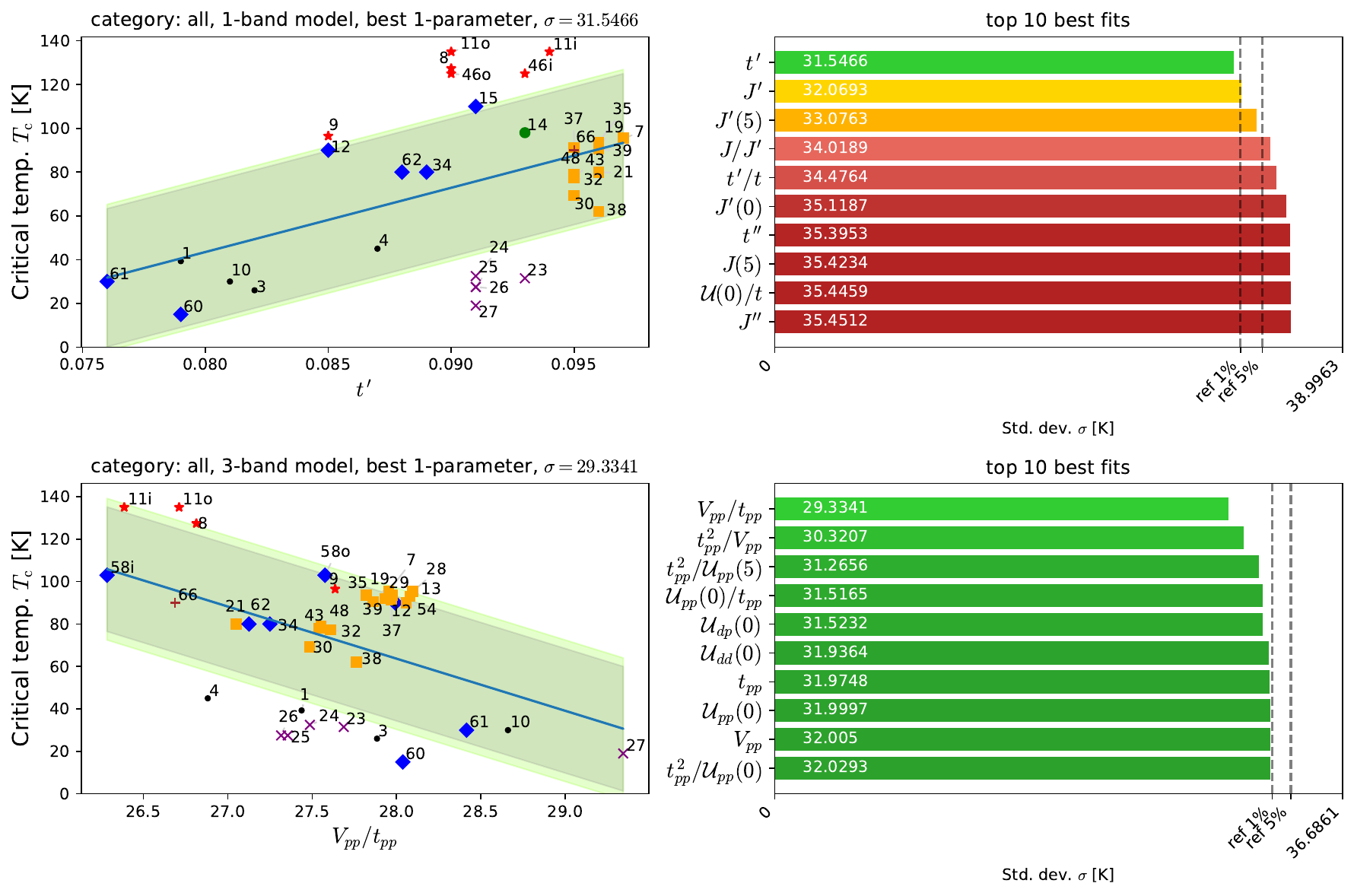}
\caption{
Correlation between individual model-parameters (as defined in the main text) and the experimentally measured $T_c$ in the entire dataset. Top panels: single-band model; bottom panels: three-band model. Left panels: the least-squares fit for the parameter that yields the smallest standard deviation (blue line is the fit, gray semi-transparent stripe is the standard deviation of the fit). Lime color stripe is the reference standard deviation $\sigma(5\%)$ - we consider fits with $\sigma<\sigma(5\%)$ statistically significant.  Data points are denoted with markers indicating the category of compound; the marker code is the same as in Figs.~\ref{fig:dft_categories} and \ref{fig:Tc_vs_category}.
Right panels: top 10 lists of the best fits. The standard deviation of the fit is shown with horizontal bars. The color of the bar indicates whether $\sigma$ is below $\sigma(1\%)$ reference value (green), between $\sigma(1\%)$ and $\sigma(5\%)$ reference values (orange), or above  $\sigma(5\%)$ reference value (red). 
The reference values are indicated with vertical dashed lines. 
The parameters $J$, $J'$ and ${\cal U}$ are stated with shorthand notation $(0)$ and $(5)$ that indicates the value of Matsubara frequency at which they are taken, expressed in eV; if no number in parentheses is stated, this indicates the bare Coulomb derived value, i.e. the limit of infinite Matsubara frequency. See Section~\ref{sec:parametrization}, Appendix~\ref{app:parameters} and Fig.~\ref{fig:tb_illustration} for explanation of parameters.
Results show that, in the single band model, $T_c$ correlates best with parameters related to the next-nearest neighbor hopping amplitude $t'$ and the onsite coupling constant. In the three-band model, $T_c$ correlates best with parameters related to the interactions on the oxygen-site orbital and the hopping between them.
}
\label{fig:best_fits_n1}
\end{figure*}

\subsection{Statistical approach}

Our general strategy will be to look for individual parameters and linear combinations of up to three parameters
that correlate well with the $T_c$; hereby we refer to the maximal $T_c$ experimentally observed in a given compound, following any amount of any kind of doping.
As the measure of correlation between a linear combination of parameters, $g = \sum_{i=1}^n c_i p_i$, and $T_c$, we will consider the standard deviation $\sigma$ that we obtain by linear regression of $T_c(g)$.
To make sure the analysis is as unbiased as possible, we consider \emph{all} possible choices of up to 3 parameters $p_i$, and for each choice we identify the coefficients $c_i$ that yield the smallest $\sigma$. In practice, we will scan through all $\{c_i\}$ (such that $\sum_{i=1}^n c_i^2 = 1$) with a moderate resolution, and then launch a Nelder-Mead minimization algorithm to pinpoint $\{c_i\}$ that minimizes $\sigma$ to a high precision. Ultimately, we can sort the linear combinations by the $\sigma$ they yield, and form a ''top list'' of linear combinations that correlate most with the $T_c$.

\begin{figure*}[ht!]
\centering
\includegraphics[width=1.0\textwidth, trim=0.0cm 0 0cm 0, clip, page=7]{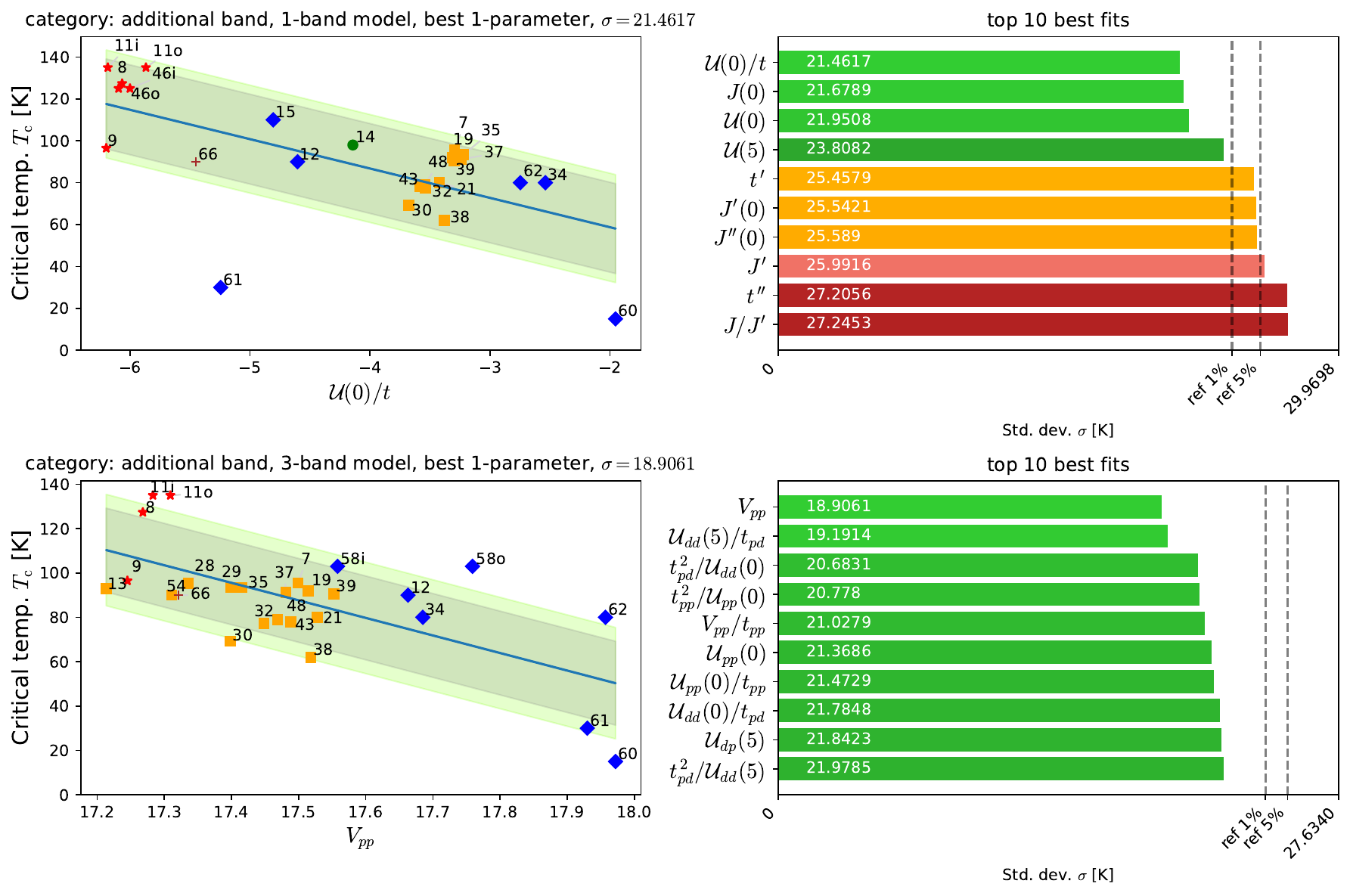}
\caption{
Same as Fig.~\ref{fig:best_fits_n1}, but for a reduced dataset, corresponding to only the categories of compounds that feature additional bands in the DFT bandstructure (columns 1-3, 5 and 6 in Fig.~\ref{fig:dft_categories}). In the one-band model, strongest correlations with the $T_c$ are observed for the interaction-to-hopping ratio, while in the three-band model, it is again the on-site interaction on the $p$-orbital that tops the chart.
}
\label{fig:additional_bands_best_fits_n1}
\end{figure*}

\begin{figure*}[ht!]
\centering
\includegraphics[width=1.0\textwidth, trim=0.0cm 10cm 0cm 0, clip, page=10]{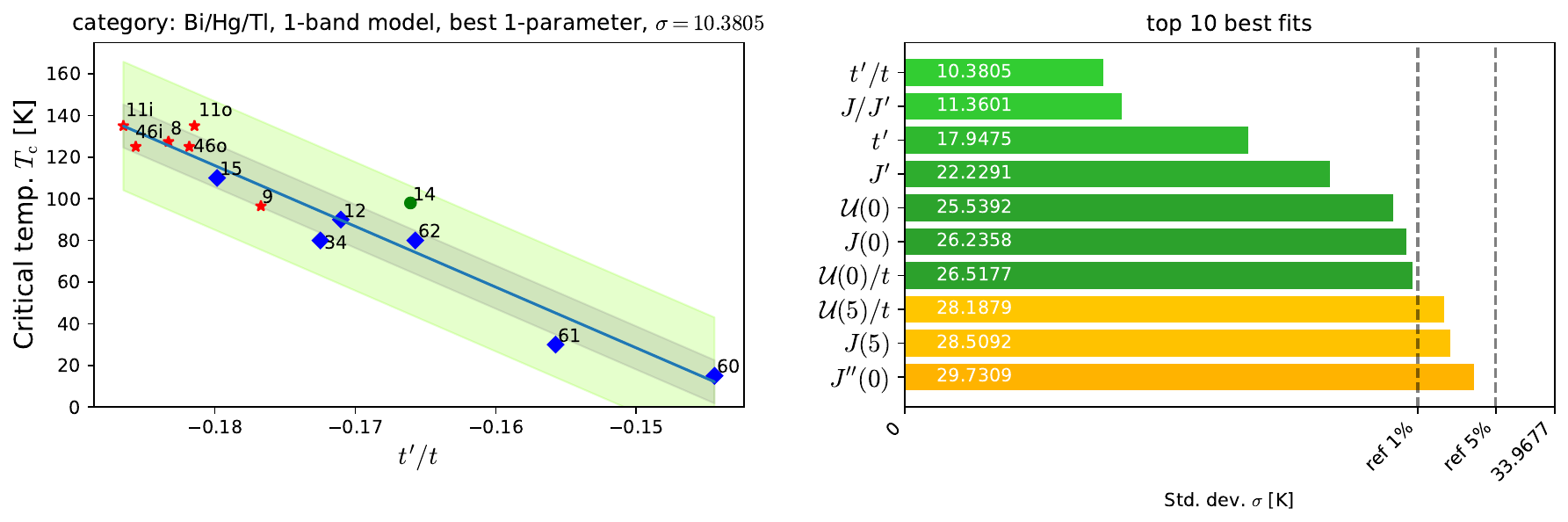}
\includegraphics[width=1.0\textwidth, trim=0.0cm 10cm 0cm 0, clip, page=11]{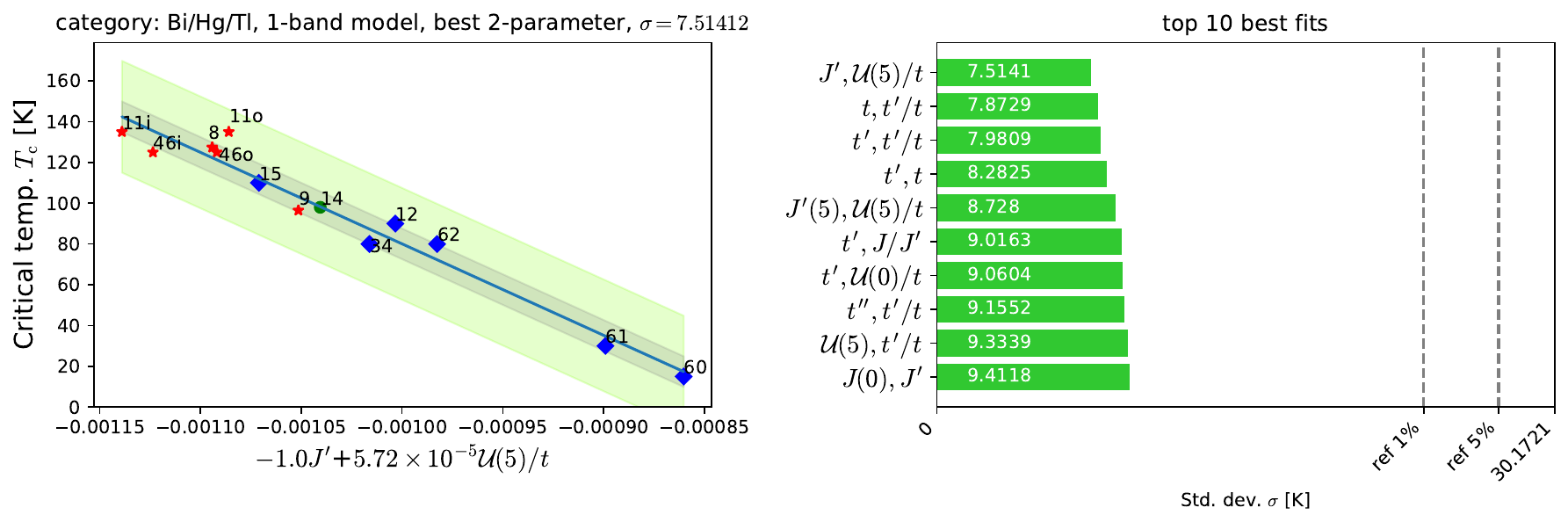}
\caption{
Same as Fig.~\ref{fig:best_fits_n1}, but for a reduced dataset corresponding to only the first 3 categories of compounds (the first 3 columns in Fig.~\ref{fig:dft_categories}), namely the compounds featuring Hg/Ba, Tl/O and Bi in the buffer layers, respectively. Furthermore, only the single-band model parameters are considered here. Top: correlations between $T_c$ and individual parameters; bottom: correlations between $T_c$ and linear combinations of two parameters. In both cases, strong correlations are observed, significantly below the reference values. As observed in previous studies\cite{Weber2012} and in ground state calculations for the Hubbard model\cite{JiangPRB2024}, the $T_c$ seems to vanish below a finite ratio $|t'/t|\sim0.1$.
}
\label{fig:additional_bands_best_fit}
\end{figure*}

The main reason we restrict ourselves to just linear combinations and a smaller number of parameters is to avoid overfitting.
Clearly, linear combinations of a larger number of parameters will generally fit better to the $T_c$, but they do not necessarily reveal more valid physical trends.
To understand whether any observed correlation is statistically significant, it is important to try and estimate the quality of fit we can expect to observe by using random numbers instead of our computed parameters.
For each section of our analysis (given by the choice of the model, the number of parameters in the linear combination, $n$, and the subset of compounds we do the linear regression for), we first perform a statistical test: we generate, 10000 times, a tuple of $n$ random numbers to use in place of the computed parameters for each of the compounds in the chosen subset, and then tune $\{c_i\}$ to minimize $\sigma$, as we do with the real data. The result is a probability distribution of $\sigma$ from which we estimate two reference values -- $\sigma(1\%)$ and $\sigma(5\%)$ -- defined as the $\sigma$ value one can expect to ``beat'' using random numbers, with the probability of 1\% and 5\%, respectively. 
In the end, the quality of the fit is not so much quantified by the absolute value of $\sigma$, but by the ratio of $\sigma$ and our reference values. We consider the cases where $\sigma<\sigma(5\%)$ statistically significant. One still needs to be careful not to overinterpret the data: if the total number of parameters is $N$, then the number of ''tries'' we get grows with $n$ as $N_\mathrm{tries}=\binom{N}{n}$, thus with larger $n$ (of course assuming $n\ll N$) we will be more likely to get results that satisfy $\sigma<\sigma(5\%)$, even with random numbers. For the given $\sigma$, $N$ and $n$, one should be able to estimate the probability of getting a better $\sigma$ result with random numbers; however, the simple formula $1-(1-P(\sigma))^{N_\mathrm{tries}}$ one would be tempted to use here [where $P(\sigma)$ is the probability of getting $\sigma$ (or better) in a single try] does not hold because the tries are not independent, but rather all draw from the same set of $N$ random numbers $p_{i=1..N}$ per compound. This probability could be evaluated numerically, but it would be computationally expensive and is beyond the scope of the current work.
Instead, we focus on the cases where we find only a few statistically significant linear combinations, or a few that clearly stand out from the rest; already with $n=3$ we usually observe a large number of linear combinations that are equally good fits to $T_c$, and such results are more difficult to interpret. Therefore, in the main text, we will ultimately restrict ourselves to analyzing only the $n\leq 2$ results.

\subsection{Most relevant examples of correlations}
\label{sec:relevant_examples}

The full results of our regression analysis are available in the Supplemental Material\cite{SMrepo}; here we discuss what we believe are the most significant findings.

We first look at the correlation of individual parameters (as defined in Section~\ref{sec:parametrization}; see Appendix~\ref{app:parameters} for the full list) with the $T_c$, across the entire dataset.
In Fig.~\ref{fig:best_fits_n1} we show the best fit to the data we have obtained using the blue line; top panels are for the Hubbard model, bottom panels are for the Emery model. The data points are labeled with ID numbers of the compounds (see Table in the Appendix~\ref{app:Tc_table}) and are shown with symbols corresponding to the categories defined in Section~\ref{sec:crystal_and_bandstructure_analysis}. The green stripe denotes $\sigma(5\%)$, and the  semi-transparent gray stripe is the $\sigma$ of the fit. For the observed correlation to be considered statistically significant, the gray stripe should be narrower than the green stripe, i.e. $\sigma<\sigma(5\%)$. On the right, we show the top lists of 10 best fits, and indicate the corresponding $\sigma$'s, as well as the reference values.

In the case of the Hubbard model, we find three parameters that yield $\sigma<\sigma(5\%)$ - all three are related to $t'$; it is $t'$ that yields the best fit, but similarly good fits are observed also with $J'$ and $J'(5)$. The latter corroborates the view that AFM interactions play an important role. Interestingly, $J'$ corresponds to the effective \emph{next-nearest}-neighbor AFM interaction, i.e. it is expected to frustrate AFM correlations on the square lattice, yet we find that it positively correlates with $T_c$. The ratio $t'/t$ also seems to be a solid predictor; in some previous studies\cite{Raimondi1996,Pavarini2001,Weber2012,Harland2016,Danilov2022,Stepanov2026} it was precisely $t'/t$ that was found to correlate with $T_c$; here we confirm this trend with a much bigger dataset, but find it might not be the best predictor, and that the correlation might not actually be statistically significant, at least not across the entire dataset (as we will see later, in certain subsets of data, correlation with $t'/t$ is indeed significant).
However, even the best fits we obtain are of poor quality. Even if the observed correlations reveal true physical trends, there must be other factors that contribute roughly $\pm 30$ K to the $T_c$.

Similar observation holds also for the Emery model. In this case, the best correlation with the data is observed for the various ratios between the hopping amplitudes and coupling constants, but primarily involving the $p$ orbitals. In fact, out of top 10 parameters (all yielding $\sigma<\sigma(1\%)$), 8 involve only ${\cal U}_{pp}$ and $t_{pp}$. This suggests that in the Emery model description, the interactions on $p$ orbitals, and the hopping amplitudes between them, are important. Even more interestingly, the frequency at which the coupling constant is taken does not appear to make much difference - the bare-Coulomb value $V_\mathrm{pp}$ appears to work equally well as ${\cal U}_{pp}(0)$ and ${\cal U}_{pp}(5)$. Certainly, the values of ${\cal U}_{pp}(0)$, ${\cal U}_{pp}(5)$ and $V_{pp}$ are expected to be correlated with one another, and it might not be surprising that they all correlate with $T_c$ similarly well. However, in Appendix~\ref{app:orientation} we show that the correlations between those parameters are not strong. This is particularly obvious when considering that ${\cal U}_{pp}(0)>{\cal U}_{dd}(0)$ in some compounds, while $V_{pp}<V_{dd}$ always - this is only possible if ${\cal U}_{\alpha\alpha}(i\nu)$ at small frequency does not quite correlate with the bare Coulomb $V_{\alpha\alpha}$.

Overall, we observe that the data points belonging to different categories of cuprates (as defined in Section~\ref{sec:crystal_and_bandstructure_analysis}), cluster on the plots, and appear to show different trends with respect to individual parameters. This suggests that it may not be possible to find a single model to describe well the $T_c$ in all of the cuprates; this motivates us to do linear regression on different categories of the cuprates separately, or group several of them together.

In particular, we observe that the compounds without any additional bands in the DFT bandstructure (purple crosses denote $T'$-structure ternary compounds with $f$-elements, black dots the rest) appear as following a separate trend from the other categories. If we exclude those compounds from the data set, we obtain more statistically significant fits with a single parameter, in both the Hubbard and the Emery model (see Fig.~\ref{fig:additional_bands_best_fits_n1}).
Now the best fit in the Hubbard model is obtained with ${\cal U}(0)/t$, and we see that stronger interactions (compared to the kinetic energy) favor superconductivity - this was already observed in a smaller dataset in Ref.~\cite{Nilsson2019}.
In the Emery model, the best fit is obtained with $V_{pp}$, while $t_{pp}^2/{\cal U}_{pp}(0)$, $V_{pp}/t_{pp}$ and ${\cal U}_{pp}(0)$ are 4th, 5th and 6th best, respectively, again indicating that the interactions on the $p$-orbitals are important for the $T_c$. 

By contrast to what we see in the case of $U$ in the Hubbard model, increasing interactions on the $p$-orbitals in the Emery model seems to reduce $T_c$ (also seen in the full dataset on Fig.~\ref{fig:best_fits_n1}).
Naively, one would expect that stronger $pp$-interactions favor the hole-doping of $p$-orbitals; indeed, it was previously observed that $T_c$ correlates with the hole-content of oxygen sites\cite{Rybicki2016,KowalskiPNAS2021}, but the trend was found to be exactly the opposite - more hole-doping on the oxygen sites meant a higher $T_c$. Thus, the effect of $p$-site interactions on the $T_c$ might be unrelated to the effect those interactions have on the occupancy of oxygen sites. The interaction on the $p$ orbitals should also be related to the charge-transfer gap $\Delta_{\mathrm{CTG}}$, which is the energy gained by an electron upon hopping from the oxygen site to the copper site - a larger interaction on the $p$-orbital should decrease the size of $\Delta_{\mathrm{CTG}}$, and thus increase the effective anti-ferromagnetic interaction $J\sim t_\mathrm{eff}^2/\Delta_{\mathrm{CTG}}$, which is expected to correlate with $T_c$\cite{Ohta1991,KowalskiPNAS2021,BacqLabreuil2025,Vadnais2026arxiv}. However, this is opposite of what we observe, thus suggesting a more complicated mechanism through which $U_{pp}$ affects the $T_c$. Another likely effect of larger interactions on the $p$-orbitals is to suppress the effective hopping between two neighboring coppers ($t_\mathrm{eff}$) - if this effect is stronger than the reduction in $\Delta_\mathrm{CTG}$, the overall effect of increasing $U_{pp}$ could well be to reduce the effective $J$, and thus the $T_c$. Indeed, some approximate treatments of the Emery model have observed the anticorrelation of $T_c$ with $U_{pp}$\cite{Esigren1998}.

We emphasize that, both in the full dataset (Fig.~\ref{fig:best_fits_n1} and the dataset restricted to compounds with additional bands (Fig.~\ref{fig:additional_bands_best_fits_n1}), the interactions on the $d$-orbitals (and even the interactions between the $d$ and $p$ orbitals) are also found to be among the best predictors. We expect all interactions to be relevant for the $T_c$ to some extent.

\begin{figure*}[ht!]
\centering
\includegraphics[width=1.0\textwidth, trim=0.0cm 0cm 0cm 10cm, clip, page=13]{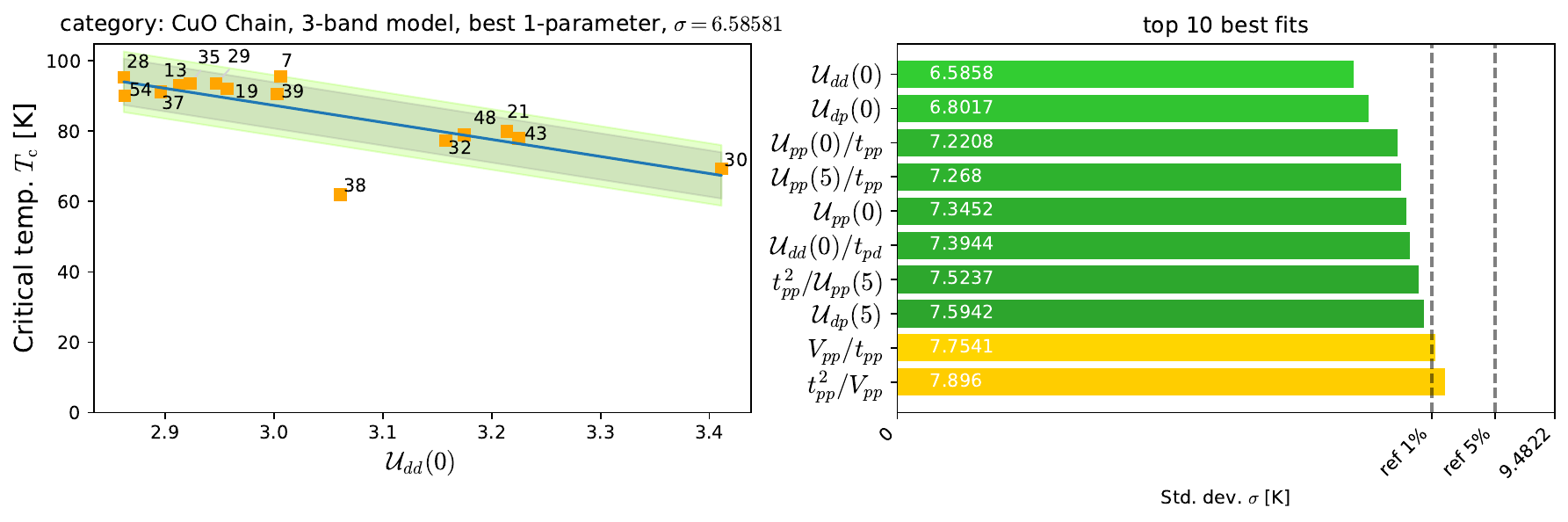}
\caption{
Analysis restricted to compounds featuring copper-oxide chains and parameters of the 3-band model. All the plotting conventions are as in previous figures. Results again point toward importance of the oxygen-site orbitals in the 3-band description.
}
\label{fig:CuO_chains_best_fit}
\end{figure*}

\begin{figure*}[ht!]
\centering
\includegraphics[width=1.0\textwidth, trim=0.0cm 0cm 0cm 10cm, clip, page=16]{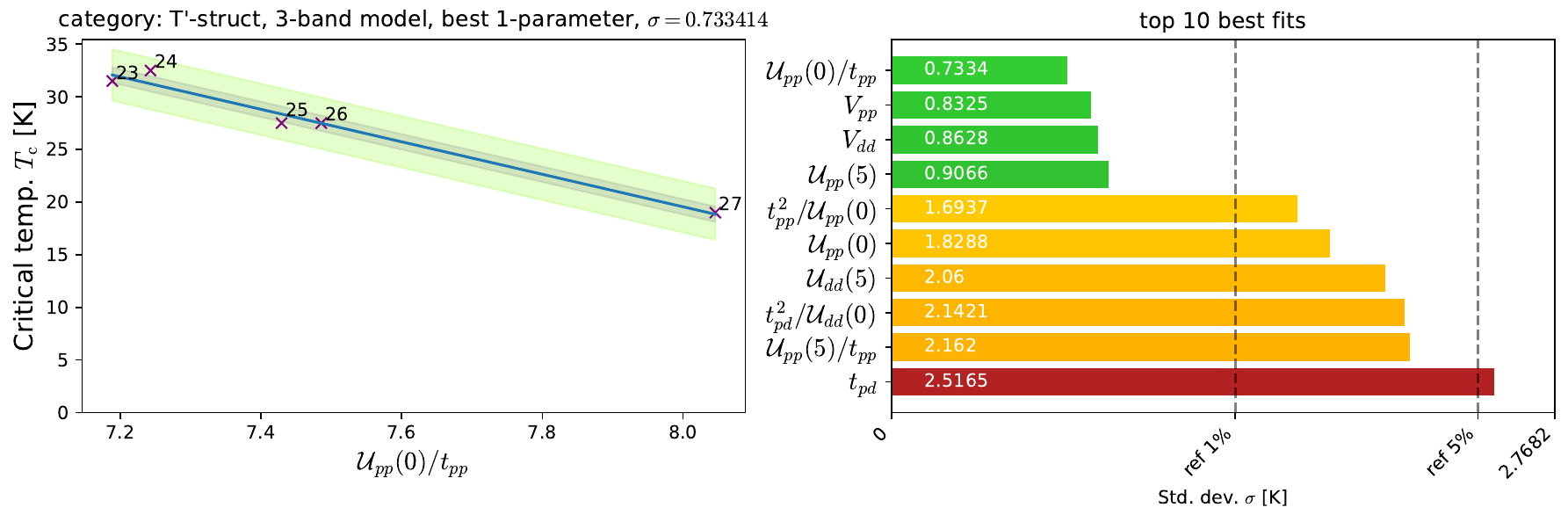}
\caption{Analysis restricted to ternary $f$-element compounds and parameters of the 3-band model. All the plotting conventions are as in previous figures.
We find that, regardless of the subset of the data, the interactions on and the hopping between the oxygen-site orbitals in the 3-band description correlate the most with the experimentally measured $T_c$.
}
\label{fig:Tp_best_fit}
\end{figure*}

\begin{figure*}[ht!]
\centering
\includegraphics[width=1.0\textwidth, trim=0.0cm 0cm 0cm 9.75cm, clip, page=1]{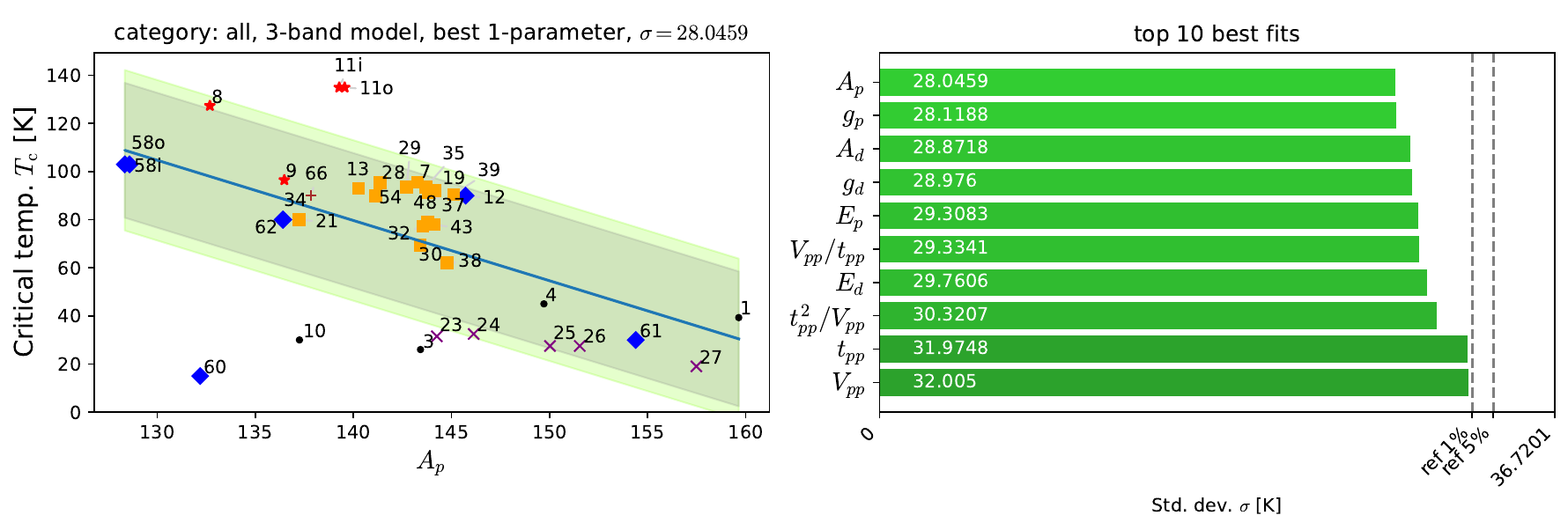}
\caption{Analysis restricted to individual parameters of the 3-band Emery-Holstein model, covering the entire dataset. All the plotting conventions are as in previous figures.
The observed scaling is better than what is found for any single parameter of the Hubbard and Emery model with instantaneous interactions.
This result suggests that a good model of cuprate $T_c$ might need to introduce retardation in the interaction, but that this can be done simply, using Holstein-model terms, with only two parameters: the coupling constant $g$ ($A=g^2$) and the boson frequency $E$, defined separately for each orbital.
}
\label{fig:all_best_plasmon_fit}
\end{figure*}

In Fig.~\ref{fig:additional_bands_best_fit} we further restrict the dataset by excluding the CuO-chain and pure Cu-layer compounds (we focus on the first three categories from the Fig.~\ref{fig:dft_categories}, i.e. the compounds with additional bands in the DFT bandstructure coming from Tl and O, Hg and Ba, and Bi atoms, respectively).
In this subset of compounds, $t'/t$ and $J'/J$ are by far the best single-parameter predictors.
It appears that $T_c$ goes to zero as $t'/t$ goes to about -0.14; in previous works with different data sets\cite{Pavarini2001, Weber2012} this was observed to happen at about $t'/t\sim-0.07$. Interestingly, these observations seem to corroborate the recent findings\cite{JiangPRB2024} that the ground state of the square lattice Hubbard model becomes superconducting only when $t'/t\lesssim-0.1$.
It is worth noting that $J'/J$ does not actually depend on the coupling strength, but physically it represents the ratio between the effective n.n. and next-n.n. AFM interactions. As both $t$ and $t'$ vary little, $t'/t$ and $J/J'$ are expected to similarly correlate with the $T_c$.
An indication of which one might actually be physically more relevant is obtained when we consider linear combinations of two parameters.
The $T_c$ in the same subset of compounds is best described by a linear combination of $J'$ and ${\cal U}(5)/t$.
The $\sigma$ of this fit is around 8K, while the reference value $\sigma(1\%)$ is around 23K; this means that it is highly unlikely that a fit of this quality is obtained with random numbers in place of our computed parameters.
It is important that this simple function depends on the effective coupling at \emph{two different} Matsubara frequencies (0 and 5eV), which indicates that a Hamiltonian description of the model might not be applicable - rather, one might need to consider the frequency dependence of the interactions as well.
We find more examples of good fits, and both $t'/t$ and the effective interactions appear often in them.
In the case of the Emery model parameters, the best predictor in this subset of compounds is a linear combination of $t_{pd}$ and $t_{pp}$ (data shown in Supplemental Material\cite{SMrepo}).

Now we focus on the other categories of compounds.
In Fig.~\ref{fig:CuO_chains_best_fit} we show the most robust trend observed in the cuprates that contain copper-oxide chains (5th category in Fig.~\ref{fig:dft_categories}).
Once again, 6 out of 10 best predictors involve only ${\cal U}_{pp}$ and $t_{pp}$.
Namely, ${\cal U}_{pp}(0)$ and ${\cal U}_{pp}(0)/t_{pp}$ anticorrelate with $T_c$ (the latter being the strongest among those 6 predictors), while ${\cal U}_{pp}(5)/t_{pp}$, ${\cal U}_{pp}(5)$, $V_{pp}/t_{pp}$, and $V_{pp}$ correlate positively with $T_c$.
However, the best two predictors are ${\cal U}_{dd}(0)$ and ${\cal U}_{dp}(0)$.
In the single-band model, the best predictor is ${\cal U}(0)/t$ (data available in the Supplemental Material\cite{SMrepo}); the correlation is somewhat weaker, but still significant. Interestingly, the trend is the opposite of what is observed in Fig~\ref{fig:additional_bands_best_fits_n1}, i.e. weaker ${\cal U}(0)/t$ means higher $T_c$. 
Overall, however, we do not obtain particularly convincing fits for this category of compounds.
This is perhaps expected, because in this category the $T_c$ varies very little.
Moreover, the additional band coming from the CuO chains crosses the Fermi level, thus it is possible that a more general model involving precisely this band is needed to describe the variation of $T_c$ in these compounds.

Finally, we focus on the ternary compounds with $f$-elements as shown in Fig.~\ref{fig:Tp_best_fit}. We have only 5 such compounds, but they all have the same crystal structure (the so-called T'-structure\cite{Das2009} that features no apex oxygens), and they only differ in the choice of the $f$-element that goes in the buffer layer - one thus expects to get good systematic trends in this subset of the data. Indeed, a linear function of the ratio ${\cal U}_{pp}(0)/t_{pp}$ predicts $T_c$ within $\pm0.73$K, i.e. with a $\sigma$ that is about 2 times smaller than $\sigma(1\%)$ for this dataset; $V_{pp}$ again yields second best $\sigma$. Both ${\cal U}_{pp}(0)/t_{pp}$ and $V_{pp}$ anticorrelate with $T_c$.

\subsection{Effective Hubbard-Holstein parameters}
\label{sec:HubbardHolstein}

Our general observation is that linear combinations of 2 and 3 parameters that yield good fits, often involve two or more interactions taken at different Matsubara frequencies.
It is possible that the effects of interaction retardation play a role, and that a good model to describe $T_c$ cannot be formulated with only instantaneous interactions.
The simplest Hamiltonian terms to introduce interaction retardation are those of the Holstein model, where electron density is coupled to a non-dispersive bosonic mode, at a frequency $\omega$, and with a coupling constant $g$, as follows
\begin{eqnarray}
 H_\mathrm{int,Holstein} &=& \omega_\alpha \sum_{i,\alpha} b^\dagger_{i,\alpha} b_{i,\alpha}\\ \nonumber
 &&+ g_\alpha \sum_{i,\alpha,\sigma} c^\dagger_{i,\alpha,\sigma}c_{i,\alpha,\sigma} (b^\dagger_{i,\alpha} + b_{i,\alpha}).
\end{eqnarray}
Here $b^\dagger/b$ are bosonic creation/annihilation operators, $i$ enumerates unit cells, $\alpha$ enumerates orbitals within the unit cell, $\sigma$ enumerates electron spin. We emphasize that the physical meaning of the boson in this model is \emph{not} that of a phonon, but rather of a charge fluctuation propagating through the rest space which is integrated out in the downfolding. Note that spin fluctuations of the Cu–O electrons, including those expected to mediate pairing, arise within the low-energy model when it is solved. The Holstein terms therefore do not double-count magnetic fluctuations; they add a retardation that the low-energy model cannot generate by itself. 

The parameters of an effective Hubbard-Holstein or Emery-Holstein model can be extracted by fitting the Matsubara on-site interaction ${\cal U}_{\alpha\alpha}(i\nu)$ obtained from cRPA to a single pole on the real axis (note that, on the real axis, positive and negative frequencies are related by symmetry)
\begin{equation}
 {\cal U}^{\mathrm{cRPA}}_{\alpha\alpha}(i\nu) \approx V_{\alpha\alpha} + A_\alpha\left(\frac{1}{i\nu-E_\alpha}-\frac{1}{i\nu+E_\alpha}\right).
\end{equation}
We have performed the fit and find that it works well for all the compounds.
The parameters for the Holstein terms in the model are then $\omega_\alpha=E_\alpha$ and $g_\alpha=\sqrt{A_\alpha}$,
while the instantaneous onsite density-density interactions have the coupling constants $V_{\alpha\alpha}$.
We then repeat the procedure for correlating parameters of the model with the experimentally measured $T_c$, just as we did in the previous section, but now we replace the interactions at particular finite frequencies (${\cal U}(0)$, ${\cal U}(5)$, etc.) with the parameters $\omega_\alpha$, $g_\alpha$ and $A_\alpha=g^2_\alpha$.
We find that $A_p$ in the 3-band model is the best single parameter to describe $T_\mathrm{c}$ in the entire dataset, see Fig.~\ref{fig:all_best_plasmon_fit}, but that all the other Holstein parameters are in top 10 predictors ($g_p$ is second best and only slightly worse).

Overall, our results suggest that an effective Emery-Holstein model with interactions on both the $d$ and the $p$-orbitals might be a good starting point to try and describe experimental $T_c$-trends.

\begin{figure*}[ht!]
\centering
\includegraphics[width=0.45\textwidth,page=1]{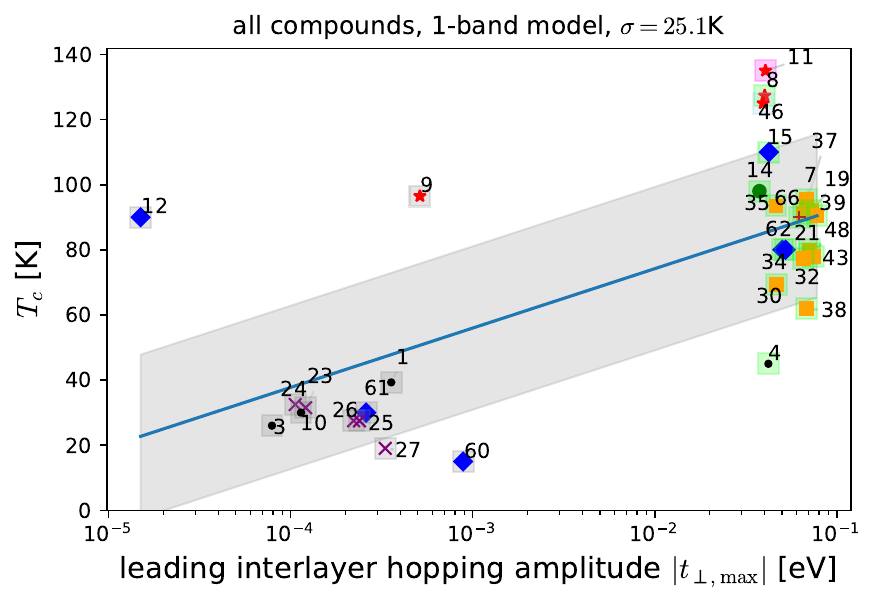}
\includegraphics[width=0.45\textwidth,page=2]{Figure_tperpmax_vs_Tc}
\includegraphics[width=0.45\textwidth,page=1]{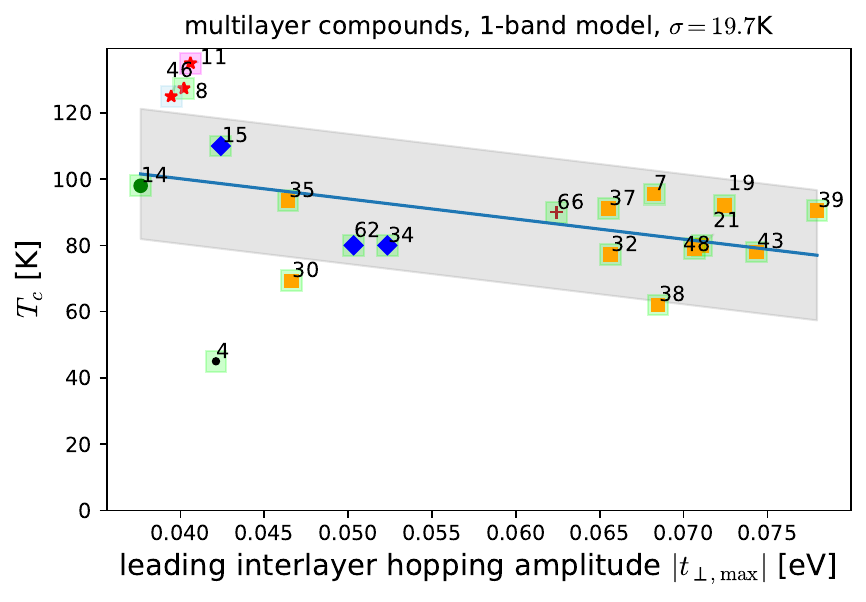}
\includegraphics[width=0.45\textwidth,page=2]{Figure_tperpmax_vs_Tc_just_multilayer}
\caption{Dependence of $T_c$ on the leading interlayer hopping $|t_{\perp,\mathrm{max}}|$ and its logarithm $\log_{10}|t_{\perp,\mathrm{max}}|$.
The symbols denote compound categories as in previous plots; in addition, here we also denote the number of CuO$_2$ layers in each compound using squares colored according to the color-scheme in Fig.~\ref{fig:Tc_vs_category} (gray 1-layer, lime 2-layer, magenta 3-layer, and skyblue 4-layer).
The value of $|t_{\perp,\mathrm{max}}|$ is strongly clustered with tiny values corresponding to single-layer compounds, and larger values corresponding to multi-layer compounds. The $T_c$ strongly correlates with $\log_{10}|t_{\perp,\mathrm{max}}|$ across the entire dataset, but within the group of multilayer compounds it anticorrelates with $|t_{\perp,\mathrm{max}}|$. Our interpretation is that interlayer hoppings are not the essential ingredient for the physical mechanism which makes $T_c$ higher in multilayer compounds.
}
\label{fig:tperpmax_vs_Tc}
\end{figure*}

\begin{figure*}[ht!]
\centering
\includegraphics[width=1.0\textwidth, trim=0.0cm 0 0cm 0, clip, page=17]{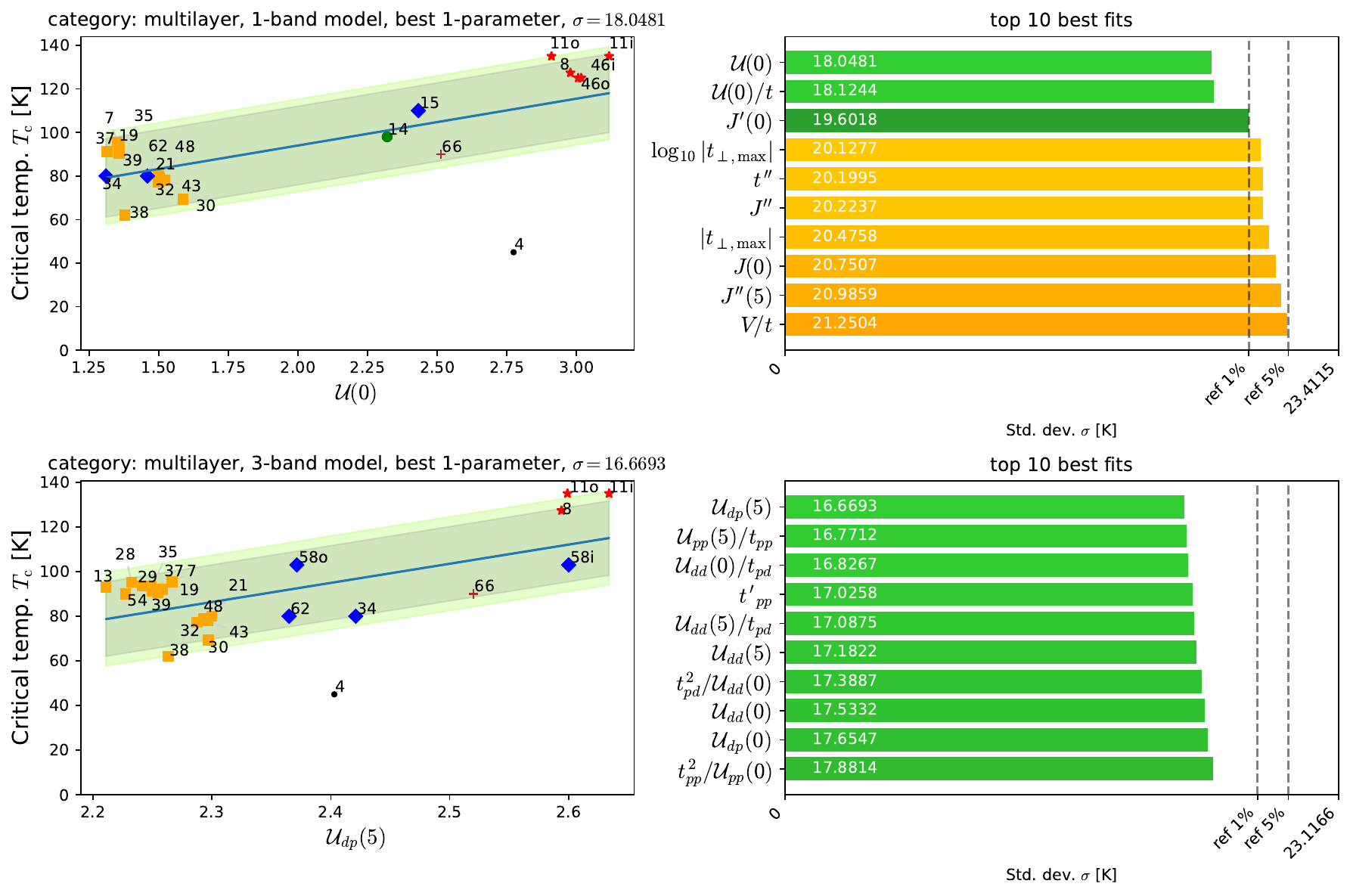}
\caption{Analysis restricted to multilayer compounds. All the plotting conventions are as in previous figures. We consider individual parameters, including the leading interlayer hopping amplitude $|t_{\perp,\mathrm{max}}|$ and its logarithm $\log_{10}|t_{\perp,\mathrm{max}}|$. 
These interlayer hopping parameters are found to not be the best predictors; instead, the best predictors consistently involve coupling constants, including ${\cal U}_{pp}$ in the 3-band model.
}
\label{fig:best_fits_tperpmax}
\end{figure*}

\subsection{Interlayer hopping}
\label{sec:interlayer}

As already stated, our goal is to parametrize the minimal effective single-layer models, keeping in mind that multilayered models are more difficult to solve. On the other hand, $T_c$ strongly correlates with the number of layers and it is important to understand if interlayer coupling is related to this apparent trend. 

Here we analyze our results with respect to interlayer hopping amplitudes. We define the leading interlayer hopping $|t_{\perp,\mathrm{max}}|$ as the largest hopping amplitude between any two Wannier orbitals belonging to separate layers of copper oxide. The results are shown in Fig.~\ref{fig:tperpmax_vs_Tc}. The immediate observation is that $|t_{\perp,\mathrm{max}}|$ values are strongly clustered. The single-layer compounds have tiny interlayer hoppings, between 0.01 and 1 meV, while multilayer compounds have sizeable interlayer hoppings between 0.04 and 0.2 eV. This is easily understood, as in multilayer compounds, the layers are more closely packed.

Sub-meV hopping amplitudes are clearly below the uncertainty of the wannierization. Any observed trends in $T_c(|t_{\perp,\mathrm{max}}|)$ within the single-layer group of compounds should not be stronger than random noise, and this is precisely what we observe.
Nevertheless, across the entire dataset, $\log_{10}|t_{\perp,\mathrm{max}}|$ correlates more with $T_c$ than any other parameter (compare Fig.~\ref{fig:tperpmax_vs_Tc} upper row with Fig.~\ref{fig:best_fits_n1}). In both models, $\sigma$ of the $T_c(\log_{10}|t_{\perp,\mathrm{max}}|)$-fit is about 25K. However, this result should not be overinterpreted. Unlike the other parameters which vary little across our dataset, $|t_{\perp,\mathrm{max}}|$ changes 4 orders of magnitude and is strongly clustered around two values - one corresponding to single-layer compounds, the other to multilayer compounds. Inspecting Fig.~\ref{fig:tperpmax_vs_Tc} we see multiple examples where $T_c$ is roughly the same between compounds in which $|t_{\perp,\mathrm{max}}|$ differs by several orders of magnitude; there are 2 single-layer compounds with $T_c\sim 100 K$ which are strong outliers to the trend, one of them having the lowest $|t_{\perp,\mathrm{max}}|$ value of all. We conclude that the trend with respect to $|t_{\perp,\mathrm{max}}|$ merely captures the fact that multilayer compounds have both a higher $T_c$ and a higher $|t_{\perp,\mathrm{max}}|$, even though interlayer hoppings themselves are not necessarily the central part of the physical mechanism behind the $T_c(N_\mathrm{layers})$ trend. In fact, none of the subsets of data studied in the previous sections show strong correlations with $|t_{\perp,\mathrm{max}}|$, and in most cases this parameter does not even show up in the top 10 list. Focusing only on the multilayer compounds (where $|t_{\perp,\mathrm{max}}|$ is sizeable enough to be reasonably well estimated from wannierization) we find that the trend with $|t_{\perp,\mathrm{max}}|$ is relatively weak, and that $T_c$ actually anticorrelates with $|t_{\perp,\mathrm{max}}|$. Comparison with other trends in this group of compounds shows that interactions are better predictors, and particularly so in the 3-band model, where $|t_{\perp,\mathrm{max}}|$ does not appear in the top 10 list. See Fig.~\ref{fig:best_fits_tperpmax}. Note that, to facilitate a fair comparison between $|t_{\perp,\mathrm{max}}|$ and $\log_{10}|t_{\perp,\mathrm{max}}|$ with other predictors, inner and outer layers in compounds with $N_\mathrm{layers}>2$ were considered here as separate data points (as we do throughout the analyses in the previous sections), and the value of $|t_{\perp,\mathrm{max}}|$ is taken the same in both layers; this leads to a slight difference between estimates of $\sigma(T_c(\log_{10}|t_{\perp,\mathrm{max}}|))$ in Fig.~\ref{fig:tperpmax_vs_Tc} and Fig.~\ref{fig:best_fits_tperpmax}, as in the former, the inner and outer layers were not treated as separate data points. The ordering of predictors by $\sigma$, however, is not affected by this choice.

Our conclusion that interlayer coupling is not directly responsible for the $T_c$-boost in multilayer compounds is in line with previous studies. In the recent Ref.~\onlinecite{BacqLabreuil2025}, the role of multilayeredness in determining the $T_c$ is rather found to be related to the existence of additional bands in the DFT bandstructure, with the orbital character coming from outside of the copper oxide planes. 
Those bands are found to limit the size of the charge-transfer gap, while their position in the bandstructure is sensitive to the number of layers.
This goes in line with our observations (made on a much larger dataset), and points towards the need for more sophisticated models, likely those that feature hybridization with some outside states (as was done in superconducting calculations in Refs.~\onlinecite{BacqLabreuil2025} and ~\onlinecite{Kitatani2020}), but not necessarily multiple layers.
The dependence of $T_c$ on the number of layers and the precise mechanism behind it must be specific to a given cuprate family; It cannot be studied across broad datasets and using minimal lattice models that we consider here.
Consider for example the following series of compounds: 9 (1-layer, $T_c=96$K), 8 (2-layer, $T_c=127$K), 11 (3-layer, $T_c=135$K) and 46 (4-layer, $T_c=125$K); they all belong to the same Hg/Ba family and form a series of increasing number of layers with $T_c$ peaking at $N_\mathrm{layers}=3$. However, already the single-layer compound (9) has a very high $T_c$, and this is likely to be strongly influenced by intraplane parameters, even if other factors are also present. In this series of compounds, having the optimal number of layers (3) boosts the $T_c$ by only about 40\%; by contrast, across our dataset the $T_c$ varies from 15 to 135K, i.e. almost an order of magnitude.

\section{Conclusions, discussion and prospects for future work}
\label{sec:conclusion}

We have compiled an extensive dataset of experimentally measured $T_c$ for various families of cuprate compounds, which we document in detail, including all relevant sources, in Appendix~\ref{app:Tc_table}.

For nearly every compound from this dataset, and solely based on the corresponding crystal structures, we have derived, from first principles, two standard microscopic models (single-band Hubbard and three-band Emery).
This was done in a fully systematic and automated manner, with no human input, treating all compounds on an equal footing.
The systematic calculation of coupling constants on this scale is unprecedented, to the best of our knowledge, and is only made possible by recent algorithmic advancements in the implementation of cRPA\cite{CoQui_GitHub,THCGW_Yeh2024}.
The result of our calculations is a large dataset of computed numerical parameters that can readily be of use for future studies of the cuprates.
We document in detail all the parameters that are needed to construct multiple versions of our models (depending on the interactions kept in the model; we provide also the DFT-level Wannier-orbital occupancies, needed to perform double-counting corrections of tight-binding parameters in a simple approximation\cite{Sheng2022}).

We have analyzed our data using linear regression, to uncover correlations between model-parameters and the experimentally measured $T_c$.
We observe that different categories of compounds, defined by crystal- and band-structure features, follow different trends.
In particular, the DFT bandstructure for a large group of compounds features additional bands near the Fermi level, with orbital character corresponding to atoms in the buffer layers; we find that, for those compounds, $T_c$ varies the most and is the highest. The $T_c$ in those compounds can be well described by a simple formula of single-band model parameters, highlighting the importance of the effective on-site interaction
(in line with previous observations\cite{Nilsson2019}), but also of its frequency dependence.
In this group of compounds, we reproduce the previously observed trend that $T_c$ correlates with $t'/t$ and that it vanishes at a finite $t'/t$\cite{Pavarini2001,Weber2012}, which is also in agreement with recent studies of the ground state in the 2D Hubbard model\cite{JiangPRB2024}. 
However, across our full dataset, we find that the parameters of the 3-band model yield better correlations with $T_c$, especially those that parametrize the effective interactions on the oxygen $p$-orbitals. This finding is reproduced in some subsets of the data that correspond to different categories of the cuprates. We propose an Emery-Holstein model to introduce interaction retardation obtained from cRPA calculation, and find that the oxygen-site coupling constant (squared) in that model is the single best (intralayer) parameter to describe $T_c$ across the entire dataset. We believe this is strong motivation to study the effect of oxygen-site interactions and interaction retardation on $T_c$ at the level of lattice models.

Our analysis rests on a sequence of approximations whose impact on the conclusions deserves to be made explicit. First, the effective interactions are evaluated within cRPA, a particular approximation to the screened Coulomb interaction that neglects vertex corrections and treats the polarization at the level of independent particle–hole excitations of the unconstrained subspace. Alternative downfolding schemes (constrained $GW$, $GW$+EDMFT, etc.\cite{Biermann2003,Nilsson2017}) are known to yield quantitatively different effective couplings and the conclusions drawn from regression analyses of renormalized interaction-like parameters should therefore be regarded as cRPA-specific. A direct comparison with results obtained from at least one alternative method, on a small subset of compounds, would help establish the robustness of our findings; this is left for future work. Second, our parametrization uses maximally localized Wannier functions to define the model orbitals. A different choice of basis (e.g. atomic-like projectors) would generally lead to different one- and two-body parameters, while reproducing the DFT low-energy bandstructure similarly well — a representation ambiguity that propagates into the regression analysis. Although the MLWF basis is the most widely used, our quantitative claims — in particular the relative ordering of our predictors — could in principle be sensitive to this choice. Third, we have restricted the effective interaction to intra-cell, density–density terms. Off-site exchange,
pair hopping, and Hund’s-like couplings are dropped from
the outset, though they could well play a role - consideration of different models including, say, Cu $d_{z^2}$ orbitals\cite{Sakakibara2010} and additional interaction terms could be relatively easily performed in the future, using the data and the algorithms we have prepared for the present study.
Finally, we use the maximal experimentally observed $T_c$ as the regression target, while parameters are derived from the stoichiometric, undoped parent compound. This implicitly assumes that the parent compound sets a $T_c$ scale that the optimal doping reaches; this is a working hypothesis, motivated by the empirical clustering\cite{Ohta1991,Pavarini2001}
of optimal $T_c$ within structural families, rather than a rigorous result. Compounds whose maximal $T_c$ is governed by
extrinsic factors — structural phase transitions involving enlargement of the unit cell, disorder introduced by doping, oxygen non-stoichiometry, or competition with charge order — may scatter for reasons unrelated to the model parameters. Moreover, the possibilities of boosting the $T_c$ by partial chemical substitution are certainly not exhausted for many of the stoichiometric compounds we consider - the highest $T_c$'s experimentally achieved \emph{so far} might not be fully representative of the physics of those compounds.

Nevertheless, the correlation between $T_c$ and the parameters of Coulomb interaction on oxygen sites appears very robust and independent of both the subset of data considered, the precise way the coupling is formulated, and even the cRPA approximation (the bare Coulomb $V_{pp}$ correlates well with $T_c$) - this finding is unlikely to be accidental. Having this in mind,
the most natural next step is a direct many-body solution of the Emery-Holstein model with the parameters tabulated here. DCA\cite{MaierRMP2005,StaarPRB2014,Maier2019}, CDMFT\cite{SakaiPRB2016,KowalskiPNAS2021}, and AFQMC\cite{ZhangKrakauer2003} are all suitable, and the relative simplicity of the on-site Holstein interaction makes the model tractable on existing platforms. A successful numerical reproduction of the experimental $T_c$-vs-$g_p$ trend across our dataset would constitute strong evidence that retardation of the local oxygen-site interaction plays a role in setting the magnitude of $T_c$. Conversely, a failure to reproduce the trend would point to physics outside the present parametrization 
and motivate richer downfolding schemes. In particular, models featuring hybridization to buffer layers states (or CuO-chains) are likely the most natural next step.

\begin{acknowledgments}
We acknowledge useful discussions with Nigel Hussey.
We acknowledge contribution by Yusuke Nomura to wannierization using RESPACK.
We acknowledge Yueqing Chang for testing our database.
Computations were performed on the PARADOX supercomputing facility (Scientific Computing Laboratory, Center for the Study of Complex
Systems, Institute of Physics Belgrade), and with Flatiron Institute's computational resources.
J.~V. acknowledges funding provided by the Institute of Physics Belgrade, through the grant by the Ministry of Science, Technological Development and Innovation of the Republic of Serbia.
J.~V. and U.~K. acknowledge funding by the European Research Council, grant ERC-2022-StG: 101076100.
M.R. acknowledges support from the Vidi ENW research programme of the Dutch Research Council (NWO) [Grant DOI: 10.61686/YDRHT18202] with File No. VI.Vidi.233.077.
The Flatiron Institute is a division of the Simons Foundation.
\end{acknowledgments}
\newpage
%\clearpage
\begin{widetext}

\appendix
\setcounter{table}{0}

\section{Experimental $T_c$ and crystallographic data}
\label{app:Tc_table}

\begin{center}
\begin{longtable}{|l|l|l|l|l|}
\caption{Experimental critical temperature data collected from the cited sources; for each $T_c$ value (in the third column) we state the doping scheme with which it was realized; otherwise, the stated $T_c$ corresponds to the stoichiometric parent compound. In the next-to-last column we provide sources for the space group of the crystal structure; in the last column we state the database ID of the crystal structure we used ("mp" stands for The Materials Project, "COD" for the Crystallography Open Database), as well as its space group. The first column is the ID we use to denote the data points on all the figures.}\label{tab:Tc}  \\
\hline \multicolumn{1}{|c|}{} & \multicolumn{1}{c|}{ \thead{{}\\\textbf{Compound}\\{}}} & \multicolumn{1}{c|}{\thead{\textbf{T$_\mathrm{c}$}\\{}}} & \multicolumn{1}{c|}{\thead{\textbf{Space group}\\\textbf{reference}}} & \multicolumn{1}{c|}{\thead{\textbf{Space group}\\\textbf{used in calculation}\\\textbf{and crystal structure ID}}} \\ \hline
\endfirsthead
\multicolumn{5}{c}%
{{\bfseries \tablename\ \thetable{} -- continued from previous page}} \\
\hline \multicolumn{1}{|c|}{} & \multicolumn{1}{c|}{\textbf{Compound}} & \multicolumn{1}{c|}{\textbf{T$_\mathrm{c}$}} & \multicolumn{1}{c|}{\thead{\textbf{Space group}\\\textbf{reference}}} & \multicolumn{1}{c|}{\thead{\textbf{Space group}\\\textbf{used in calculation}\\\textbf{and crystal structure ID}}} \\ \endhead

\hline \multicolumn{5}{r}{{Continued on next page}} \\
\endfoot

\hline
\endlastfoot
&&&&\\
\quad 1 \quad &\quad  La$_2$CuO$_4$ &\quad  La$_{2-x}$Sr$_x$CuO$_{4-\delta}$ \quad  & \quad I4/mmm \cite{Das2009}& \quad \quad \quad  I4/mmm  \\
&&\quad  \quad  \quad  \quad 38 K \cite{trofimov1994growth}, $x=0.15$, $\delta=0$ &&\quad (mp-19735)\\
&&\quad  \quad  \quad  \quad 39.3 K \cite{tarascon1987superconductivity}, $x=0.15$, $\delta$= N/A  & &\\
&&\quad  \quad  \quad  \quad 37.5 K \cite{chaillout1989crystal}, $x=0$, \mbox{$\delta=-0.032$}   &\quad Cmca and Acam \cite{chaillout1989crystal} \quad  &  \\
&  &\quad  La$_{2-x}$Ba$_x$CuO$_{4}$ \quad  & & \quad \quad \quad    \\
&&\quad  \quad  \quad  \quad 32 K \cite{Guguchia2023}, $x=0.1$ &&\quad \\
&  &\quad  La$_{2-x}$Ce$_x$CuO$_{4}$ \quad  & & \quad \quad \quad    \\
&&\quad  \quad  \quad  \quad 30 K \cite{naito2000superconducting}, $x=0.15$ &&\quad \\
&&&&\\
&&&&\\
\hline
&&&&\\
\quad 3  \quad &\quad Ca$_2$CuO$_2$Cl$_2$  &\quad \quad \quad \quad 26 K \cite{PhysRevB.51.8434}  & \quad I4/mmm \cite{PhysRevB.51.8434} &\quad \quad  \quad I4/mmm   \\
&&&&\quad \quad (mp-23143)\\
&&&&\\
\hline
&&&&\\
\quad 4 \quad& \quad La$_2$CaCu$_2$O$_6$   &\quad La$_{2-x}$Ca$_{1+y}$Cu$_2$O$_{6+\delta}$ &\quad  I4/mmm \cite{fuertes1990oxygen} & \quad \quad \quad I4/mmm \\
&&\quad \quad \quad \quad 45 K \cite{ishii1991single}, $x=0.13$, $y=0.30$, $\delta=0$ &&\quad \quad(1006107-COD)\\
&&\quad \quad \quad \quad 45 K \cite{fuertes1990oxygen}, $x=0$, $y=0$, $\delta=0.037$ & &\\
&&&&  \\
&&&&  \\
\hline
&&&&\\
\quad 7 & \quad YBa$_2$Cu$_3$O$_7$ &\quad  YBa$_2$Cu$_3$O$_{7-\delta}$ && \\
&&\quad \quad \quad \quad 93 K \cite{Pavarini2001}, $\delta=0$  &&\\
&&\quad \quad \quad \quad 87$\pm$0.5 K \cite{schneemeyer1987superconductivity}, $\delta=0$& \quad Pmmm \cite{beno1987structure} &\quad \quad  Pmmm\\
&& \quad \quad \quad \quad 92.5$\pm$3 K \cite{beno1987structure},  $\delta$= N/A&  &\quad \quad  (mp-20674)  \\
&&  &  & \\
\hline
&&&&\\
\quad 8& \quad  HgBa$_2$CaCu$_2$O$_6$ \quad  & \quad  HgBa$_2$CaCu$_2$O$_{6+\delta}$   & &  \\
& &\quad \quad \quad \quad\mbox{127 K \cite{loureiro1993synthesis}}, $\delta=0.22$ &\quad  P4/mmm \cite{loureiro1993synthesis} &\quad  \quad  P4/mmm \\
&&\quad \quad \quad \quad 120 K \cite{xie2000fabrication}, $\delta=0$&& \quad \quad (mp-6879) \\
&&\quad \quad \quad \quad 127.39 K \cite{Pavarini2001}, $\delta=0$&&  \\
&&&& \\
\hline
&&&& \\
\quad 9 & \quad HgBa$_2$CuO$_4$  &\quad Hg$_{1-x}$Cu$_y$Ba$_2$CuO$_{4+\delta}$ & &   \\
&&\quad \quad \quad \quad 94 K \cite{putilin1993superconductivity,jha1996phonon}, \quad  $x=0$, $y=0$, $\delta$=N/A &&\\
&&\quad \quad \quad \quad   96 $\pm$ 0.5 K \cite{PhysRevB.55.12776}, $x=0.1$, $y=0.05$, $\delta$=N/A   &&\quad \quad  (mp-6562) \\
&&& & \\
\hline
&& && \\
\quad 10 \quad&\quad  Sr$_2$CuO$_2$Cl$_2$   &\quad Sr$_2$CuO$_{2+\delta}$Cl$_{2-y}$  & & \\
&&\quad \quad \quad \quad 30 K \cite{yang2006symmetry}, $y$ = N/A, $\delta$ = N/A &&\\
&   &\quad \quad \quad \quad 30 K \cite{liu2005high}, $y=0.8$, $\delta$=N/A & \quad I4/mmm \cite{liu2005high} &\quad \quad   I4/mmm \\
& &&&\quad \quad  (mp-23102)\\
& &&&\\
\hline
&&&& \\
\quad 11  \quad&\quad  HgBa$_2$Ca$_2$Cu$_3$O$_8$\quad    & \quad  HgBa$_2$Ca$_2$Cu$_3$O$_{8+\delta }$ & &  \\
&&\quad \quad \quad \quad 135 K \cite{fukuoka1997dependence}, $\delta=0.29$ &&\\
&&\quad \quad \quad \quad 135 K \cite{Wagner1995}, $\delta=0.18$ &&\\
&&\quad \quad \quad \quad 135 K \cite{jha2000lattice}, $\delta$ = N/A &\quad P4/mmm \cite{jha2000lattice}&\quad \quad P4/mmm \\
&&&&\quad \quad (mp-22601) \\
&&&& \\
%&&&& \\
\hline
&&&& \\
\quad 12  \quad&\quad  Tl$_2$Ba$_2$CuO$_6$  &\quad \quad \quad \quad 90 K \cite{hermann1993single,strange1989comparison} & &\\
&&\quad \quad \quad \quad 82 $\pm$ 13 K \cite{tsvetkov1998global} &\quad  I4/mmm \cite{tsvetkov1998global} &\quad \quad    I4/mmm   \\
&& & &\quad \quad (mp-550722)\\
&&&& \\
\hline
&&& & \\
\quad 13 \quad& \quad LaBa$_2$Cu$_3$O$_7$   &\quad LaBa$_2$Cu$_3$O$_{7-\delta}$ & & \\
&   &\quad \quad \quad \quad 79 K \cite{li1987laba2cu3o7}, $\delta$= N/A & & \\
&   &\quad \quad \quad \quad 74 K \cite{wada1988high}, $\delta=0$ & & \\
&  &\quad \quad \quad \quad 93 K \cite{PhysRevB.40.6771}, $\delta=0.15$  & \quad  Pmmm \cite{PhysRevB.40.6771}&\quad \quad   Pmmm \\
&&  && \quad \quad    (mp-622210)  \\
&&  &&   \\
\hline
&& &&\\
\quad 14 & \quad Bi$_2$Sr$_2$CaCu$_2$O$_8$  &\quad  Bi$_2$Sr$_2$CaCu$_2$O$_{8+\delta}$& & \\
&  &\quad \quad \quad \quad  80 K \cite{hewat1988superstructure}, $\delta=0$ &&\\
&  & \quad \quad \quad \quad  83 K \cite{manzke1989superconducting}, $\delta$=0 &&\\
&  & \quad \quad \quad \quad 85 K \cite{cooper1990direct}, $\delta$=0 &&\\
&  &\quad \quad \quad \quad 90 K \cite{PhysRevB.53.2245}, $\delta=0$ &&\\
&  &\quad \quad \quad \quad 66 K \cite{PhysRevX.11.031068}, $\delta$=N/A&&\\
&&\quad \quad  \quad \quad 86 K \cite{PhysRevB.56.8426}, $\delta=0.13$ &\quad  I4/mmm  \cite{PhysRevB.56.8426} &\quad \quad I4/mmm   \\
&&\quad Bi$_{1.4}$Pb$_{0.7}$Sr$_{2}$Ca$_{0.97}$Y$_{0.03}$Cu$_{2}$O$_{8+\delta}$ && \quad \quad (mp-555855)\\
&&\quad \quad  \quad \quad  98 K \cite{PhysRevB.79.064507}, $\delta$=N/A&&\\
&& && \\
\hline
&& && \\
\quad 15 &\quad  Tl$_2$Ba$_2$CaCu$_2$O$_8$   &\quad \quad \quad \quad  110 K \cite{molchanov1994structure} &\quad  I4/mmm \cite{molchanov1994structure} & \quad \quad I4/mmm   \\
&&  && \quad(mp-6885) \\
&&  &&  \\
&&  &&  \\
\hline
&& && \\
\quad 16 &\quad  Bi$_2$Sr$_2$Ca$_2$Cu$_3$O$_{10}$  &\quad \quad \quad \quad  110 K \cite{lejay1989superconductivity}  & \quad \quad \quad \quad N/A & \quad \quad \quad \quad   \\
&&\quad \quad \quad \quad 109 K \cite{giannini2003growth}   & \quad \quad  \quad  A2${aa}$ \cite{giannini2003growth}&\quad \quad    \\
&& \quad (Bi$_{0.8}$Pb$_{0.2}$)$_2$Ca$_2$Cu$_3$Sr$_2$ O$_{\delta}$   &\quad\quad I4/mmm  \cite{zhu1989structure} & \quad \quad\quad I4/mmm \\
&& \quad \quad \quad \quad 111 K \cite{zhu1989structure}, $\delta$=N/A & \quad\quad (1540770-COD) &  \quad\quad\quad (mp-1209015)\\
%\hline
&& \quad  Bi$_{2.1}$Sr$_{1.9}$Ca$_2$Cu$_3$O$_{10+\delta}$ \quad  & & \\
&& \quad  \quad  \quad  \quad 105 K \cite{watanabe2003structural}, $\delta$ = N/A & &\\
&&&&\\
\hline
&&&&\\
\quad 19&\quad ErBa$_2$Cu$_3$O$_7$ &\quad \quad \quad \quad 91$\pm$ 1 K \cite{chattopadhyay1988evidence} &\quad N/A &\quad \quad Pmmm \\
&&&&\quad \quad (mp-622110)\\
&&&&\\
\hline
&&&&\\
\quad 21& \quad YBa$_2$Cu$_4$O$_8$ &\quad \quad \quad \quad 80 K \cite{lightfoot1991redetermination} &\quad  Ammm \cite{lightfoot1991redetermination} &\quad \quad  Ammm \\
&&&&\quad \quad (\mbox{COD ID: 1000031}) \\
&&&&\\
\hline
&&&&\\
\quad 23& \quad Pr$_2$CuO$_4$ &\quad \quad \quad \quad 31.5 K \cite{PhysRevB.79.100508}& \quad N/A & \quad \quad I4/mmm \\
&&&&\quad \quad (mp-4181)\\
&& && \\
\hline
&& && \\
\quad 24&\quad  Nd$_2$CuO$_4$ &\quad \quad \quad \quad 32.5 K \cite{PhysRevB.79.100508} &\quad  I4/mmm \cite{wilhelm2000pressure} &\quad \quad I4/mmm	  \\
&& &&\quad \quad (mp-4158)\\
&&&&\\
\hline
&&&&\\
\quad 25&\quad  Sm$_2$CuO$_4$ &\quad \quad \quad \quad 27.5 K \cite{PhysRevB.79.100508}&\quad N/A &\quad \quad I4/mmm \\
&&&&\quad \quad (mp-4210)\\
%&&&&\\
\hline
&&&&\\
\quad 26&\quad  Eu$_2$CuO$_4$ &\quad \quad \quad \quad 27.5 K \cite{PhysRevB.79.100508}& \quad N/A &\quad \quad  I4/mmm  \\
&&&&\quad \quad (mp-22306)\\
&&&&\\
\hline
&&&&\\
\quad 27&\quad  Gd$_2$CuO$_4$ &\quad \quad \quad \quad  19.0 K \cite{PhysRevB.79.100508}&\quad N/A &\quad \quad I4/mmm  \\
&&&&\quad \quad (mp-4860)\\
&&&&\\
\hline
&&&&\\
\quad 28&\quad NdBa$_2$Cu$_3$O$_{7}$  &\quad NdBa$_2$Cu$_3$O$_{7-\delta}$ & \quad N/A &\quad \quad Pmmm \\
&& \quad \quad \quad \quad 95.3 K \cite{PhysRevB.36.226}, $\delta=$ N/A&&\quad \quad (mp-22719)\\
&&&&\\
\hline
&&&&\\
\quad 29&\quad SmBa$_2$Cu$_3$O$_{7}$ &\quad SmBa$_2$Cu$_3$O$_{7-\delta}$ &\quad N/A &\quad \quad  Pmmm \\
&&\quad \quad \quad \quad 93.5 K \cite{PhysRevB.36.226}, $\delta=$ N/A&&\quad \quad (mp-21451)\\
&&&&\\
\hline
&&&&\\
\quad 30&\quad EuBa$_2$Cu$_4$O$_{8}$ &\quad \quad \quad \quad 68.9 $\pm$ 0.4 K \cite{PhysRevB.39.7347}&\quad N/A  &\quad \quad Cmmm \\
&&&&\quad \quad (mp-1214709)\\
&&&&\\
\hline
&&&&\\
\quad 31&\quad GdBa$_2$Cu$_4$O$_{8}$ &\quad \quad \quad \quad 73.4 $\pm$ 1.2 K \cite{PhysRevB.39.7347}&\quad N/A  &\quad \quad Cmmm \\
&&&&\quad \quad (mp-1214720)\\
&&&&\\
\hline
&&&&\\
\quad 32&\quad DyBa$_2$Cu$_4$O$_{8}$ & \quad \quad \quad \quad 77.2 $\pm$ 0.1 K \cite{PhysRevB.39.7347}&\quad N/A  &\quad \quad Cmmm \\
&&&&\quad \quad (mp-6691) \\
&&&&\\
\hline
&&&&\\
\quad 34&\quad TlBa$_2$CaCu$_2$O$_{7}$ & \quad \quad \quad \quad 80 K \cite{nakajima1990superconductivity}&\quad N/A &\quad \quad P4/mmm \\
&&&&\quad \quad (mp-632802)\\
&&&&\\
\hline
&&&&\\
\quad 35&\quad EuBa$_2$Cu$_3$O$_7$ & \quad \quad \quad \quad 93.5 K \cite{PhysRevB.40.6948}&\quad N/A  &\quad \quad Pmmm \\
&&&&\quad \quad(mp-622211)\\
&&&& \\
\hline
&&&& \\
\quad 36&\quad GdBa$_2$Cu$_3$O$_7$& \quad \quad \quad \quad  95 K \cite{PhysRevB.40.6948}&\quad N/A &\quad \quad Pmmm \\
&&&& \quad \quad (mp-19813)\\
&&&&\\
\hline
&&&&\\
\quad 37&\quad DyBa$_2$Cu$_3$O$_7$& \quad \quad \quad \quad  91.2 K \cite{kolodziejczyk1993electronic}&\quad N/A &\quad \quad Pmmm \\
&&&&\quad \quad (mp-622105)\\
&&&&  \\
\hline
&&&&  \\
\quad 38&\quad HoBa$_2$Cu$_3$O$_7$& \quad HoBa$_2$Cu$_3$O$_{7-\delta}$ \quad \quad \quad &\quad N/A  &\quad \quad Pmmm \\
&&\quad\quad\quad\quad 62 K \cite{solovjov2014fluctuation}, $\delta=$ N/A  &&\quad \quad (mp-6616)\\
&&&&\\
\hline
&&&&\\
\quad 39&\quad TmBa$_2$Cu$_3$O$_7$&\quad \quad \quad \quad  90.5 K \cite{ishigaki1987crystal}&\quad N/A &\quad \quad Pmmm \\
&&&&\quad \quad (mp-622108)\\
\hline
&&&&\\
\quad 43   \quad &\quad ErBa$_2$Cu$_4$O$_8$  &\quad \quad \quad \quad 78 K \cite{van1990normal}  &\quad \quad N/A &\quad \quad  \quad Cmmm   \\
&&&&\quad \quad (mp-6583)\\
\hline
&&&& \\
\quad 46&\quad HgBa$_2$Ca$_3$Cu$_4$O$_{10}$ & \quad  HgBa$_2$Ca$_3$Cu$_4$O$_{10+\delta}$  &\quad \quad N/A & \quad \quad P4/mmm \\
&&\quad  \quad  \quad  \quad  125 K \cite{kim2000nature}, $\delta$= N/A  &  & \quad \quad (mp-1228579) \\
&&  & & \\
\hline
&&&& \\
\quad 48&\quad HoBa$_2$Cu$_4$O$_8$& \quad \quad \quad 79 K \cite{rubio2001large} &\quad \quad  N/A &\quad  Cmmm \\
&&  & &\quad  (mp-6205)\\
&&  & & \\
\hline
&&&& \\
\quad 54&\quad PrBa$_2$Cu$_3$O$_7$ & \quad \quad \quad 90 K \cite{blackstead1995superconductivity} &\quad \quad N/A  &\quad  Pmmm \\
&&  & &\quad (mp-20936) \\
&&&&\\
\hline
&&&& \\
\quad 58&\quad TlBaSrCa$_2$Cu$_3$O$_9$ & \quad \quad TlBaSrCa$_2$Cu$_3$O$_{9-\delta}$  & \quad \quad P4/mmm  \cite{martin1992116}& \quad \quad  P1m1 \\
&&\quad \quad \quad \quad 103 K \cite{martin1992116}, $\delta=$ N/A  &  &\quad \quad (mp-1227794)\\
&&&&\\
\hline
&&&& \\
\quad 59&\quad TlBa$_2$Ca$_3$Cu$_4$O$_{11}$ & \quad \quad TlBa$_2$Ca$_3$Cu$_4$O$_{11-\delta}$ & \quad \quad P4/mmm \cite{zhang1997crystal} & \quad \quad Pmm2 \\
&&\quad \quad \quad \quad 128 K \cite{zhang1997crystal}, $\delta$= N/A&  &  \quad \quad (mp-1228589) \\
&&&&\\
\hline
&&&& \\
\quad 60&\quad TlBa$_2$CuO$_5$&  \quad \quad TlBa$_2$CuO$_{5-\delta}$ & \quad \quad N/A  & \quad \quad  P4/mmm  \\
&&\quad \quad \quad 15 K \cite{shi1989superconductivity}, $\delta$= N/A  &  & \quad \quad (mp-20942)\\
&&&&\\
\hline
&&&& \\
\quad 61&\quad TlLaSrCuO$_5$  & \quad \quad 30 K \cite{nagashima1990synthesis} &\quad \quad P4mm \cite{nagashima1990synthesis} & \quad \quad P4mm  \\
&&  & & \quad \quad (mp-1218170)\\
&&&&\\
\hline
&&&& \\
\quad 62&\quad TlSr$_2$CaCu$_2$O$_7$&\quad \quad  (Tl$_{0.5}$Pb$_{0.5}$)Sr$_2$CaCu$_2$O$_7$& \quad \quad P4/mmm \cite{liu1991enhancement}& \quad \quad P4/mmm   \\
&&\quad \quad \quad 80 K \cite{liu1991enhancement}  &  &\quad \quad  (mp-20824) \\
\hline
&&&& \\
\quad 66&\quad YBa$_2$Cu$_3$O$_6$&\quad \quad  YBa$_2$Cu$_3$O$_{6+\delta}$ &\quad \quad  N/A  & \quad P4/mmm \\
&&\quad \quad \quad 90 K \cite{rezlescu1995model}, $0.8 \leq \delta \leq 1$& &\quad (mp-22215) \\
\end{longtable}
\end{center}

\end{widetext}

\begin{figure*}[ht!]
\centering
\includegraphics[width=1.0\textwidth]{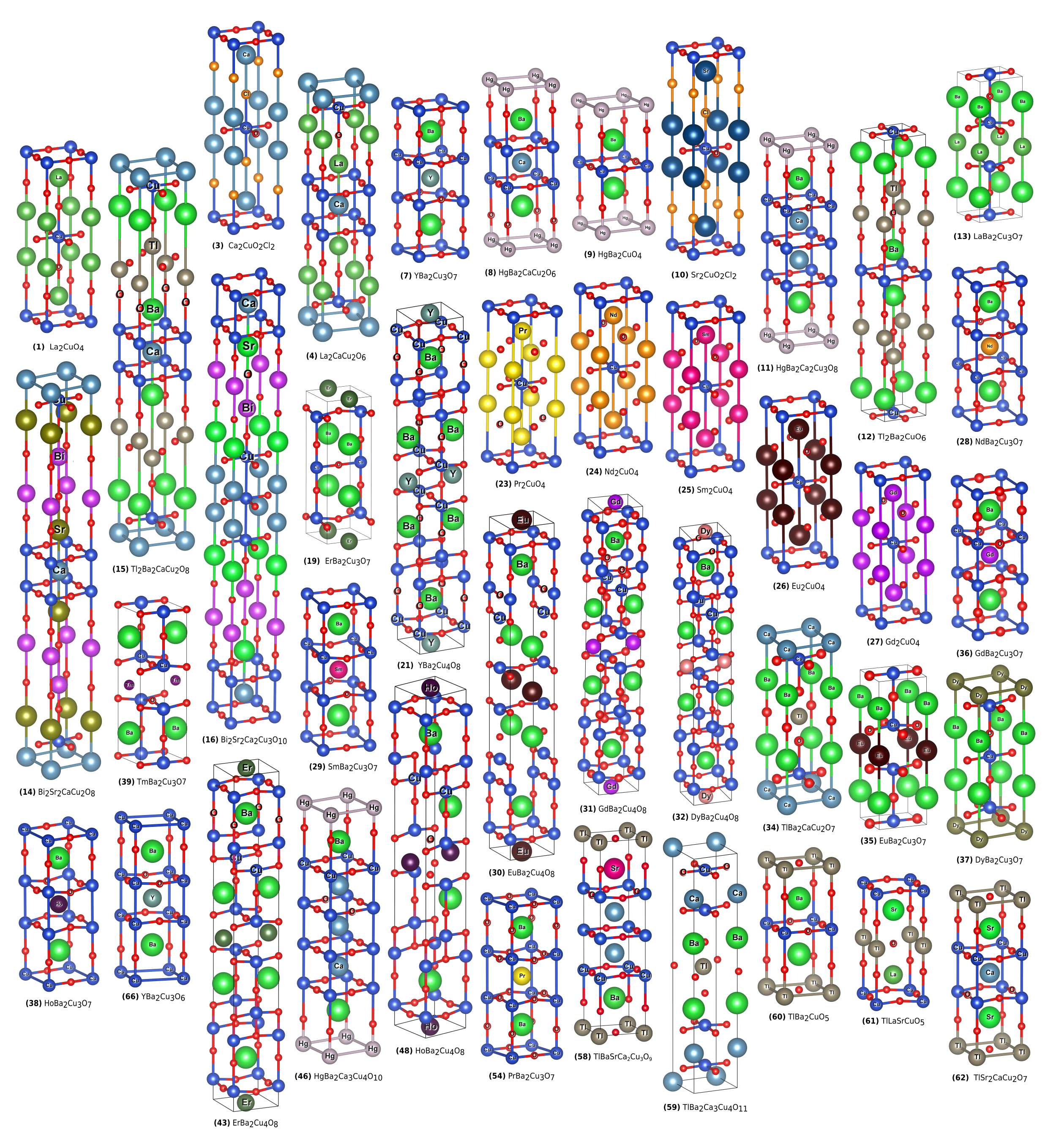}
\caption{Conventional unit cells of cuprate compounds considered in this work, illustrated using VESTA\cite{Momma2011}.
}
\label{fig:collage}
\end{figure*}

\appendix
\setcounter{section}{1}

\section{DFT calculations and wannierization}
\label{app:dft_and_wannierization}

Here we outline the steps we take to prepare inputs for the DFT calculation and the wannierization for each compound.

We start by selecting the crystal structure for the given compound, as explained in the main text. We make sure that there is no enlargement of the unit cell, i.e. that there is a single copper atom and two oxygen atoms per CuO$_2$ layer per primitive unit cell.
We then rotate the lattice vectors to place copper-oxide layers in the $xy$ plane and to orient the copper-oxygen bonds along the $x$ and $y$ directions.  
This step is needed to make sure that, when we initialize the wannierization with the copper $d_{x^2-y^2}$ and oxygen $p_x$ and $p_y$ orbitals, they are appropriately oriented within the crystal structure.
Some symmetries of the lattice are evident in the coordinates of the lattice vectors, where certain coordinates have the same numerical value.
However, for some of the lattice structures we used, some symmetries were not perfectly satisfied, i.e. some coordinates were almost the same, but differed slightly. 
When the difference between any two coordinates was present only beyond the fourth decimal digit, we have made those coordinates exactly the same, so that in the DFT calculation additional symmetries can be used to speed up the calculation.
The DFT calculation is then performed, using the primitive unit cell to avoid any possible breaking of symmetry.
The parameters of the DFT self-consistent calculation were set the same for all compounds.
Wavefunction cutoff was set at 80 Ry, and Gaussian smearing was set at 0.02 Ry.
We have done the calculations (as well as subsequent wannierization and cRPA steps) with 6$\times$6$\times$6 and 8$\times$8$\times$8 k-points and confirm solid level of convergence, and we then analyze and present the 8$\times$8$\times$8 results.

\begin{figure}[ht!]
\centering
\includegraphics[width=0.45\textwidth,page=1]{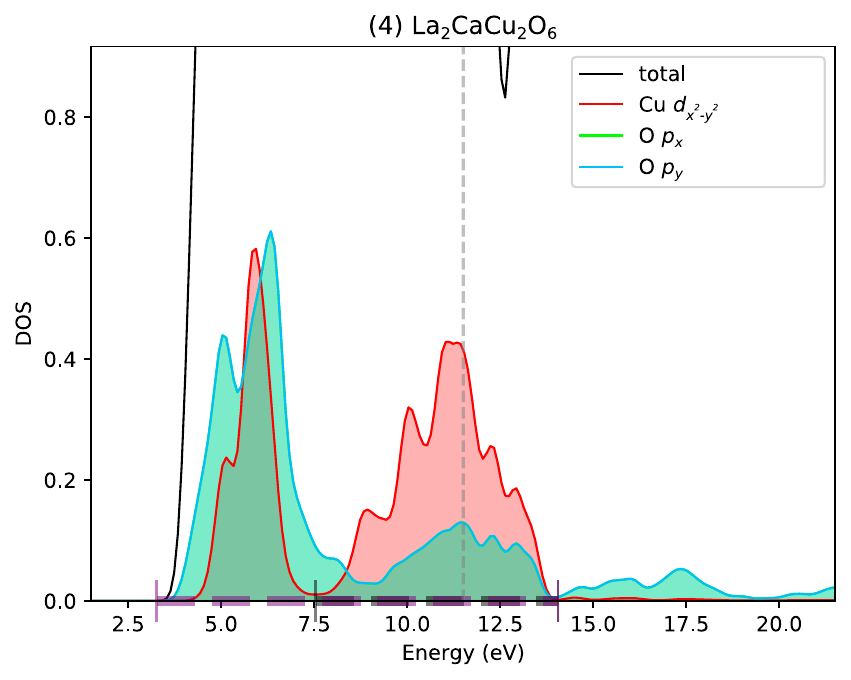}
\caption{Example of the partial density of states computed for the atomic orbitals, Cu 3$d_{x^2-y^2}$ and O 2$p_{x/y}$, of the copper and oxygen atoms that form the copper-oxide planes, and the energy windows for wannierization determined from it. Data shown is for compound 4. The vertical gray dashed line is Fermi level. The horizontal dashed lines with vertical lines as endpoints are the energy windows: purple for the 3-band model, gray for the 1-band model.
}
\label{fig:partial_dos}
\end{figure}

To initialize the wannierization, we needed to identify the copper and oxygen atoms that form the copper-oxide layers, as well as enumerate and label the layers as inner or outer (the latter was relevant where more than 2 layers per primitive unit cell are present).
This is the only step in the procedure that was done manually, by visually inspecting the crystal structure in VESTA\cite{Momma2011} and filling in the corresponding input files. 

Other than the initial guess for the Wannier orbitals, the main input parameter for the wannierization is the energy window.
We have set the energy window automatically, by determining the extent in energy of the partial density of states for the atomic, initial-guess orbitals, as obtained from DFT. A subtlety in this step was that the wannierization was done based on DFT results with ONCV+DOJO pseudopotentials, while the partial densities of states we were only able to compute using SSSP pseudopotentials. We have checked that the DFT bandstructures obtained with two different sets of pseudopotentials were very similar, but they did often differ in terms of the overall energy shift, i.e. the Fermi energy was different. When setting up the energy window for wannierization, we took this into account, and shifted the energy window obtained from DFT(SSSP) results accordingly for the wannierization based on DFT(ONCV+DOJO) results.
An example of the partial densities of states and the corresponding energy windows is given in Fig.~\ref{fig:partial_dos}, while full results are available in Supplemental Material\cite{SMrepo}. In several cases where 1-band wannierization using \texttt{Wannier90} failed, we have tried to change slightly the energy window by hand; in most of those cases this did not help converge the MLWF's, but in very few it did allow us to obtain the results (although the results obtained were practically the same as the ones obtained from RESPACK, where we did not have to adjust the windows manually; therefore the manual adjustment did not introduce any bias in our data).
In some of the cases we also found that using frozen windows helps converge the MLWF's, but we opted not to use those data, as a way of ensuring equal treatment of all the compounds.

\begin{figure}[ht!]
\centering
\includegraphics[width=\columnwidth,page=5]{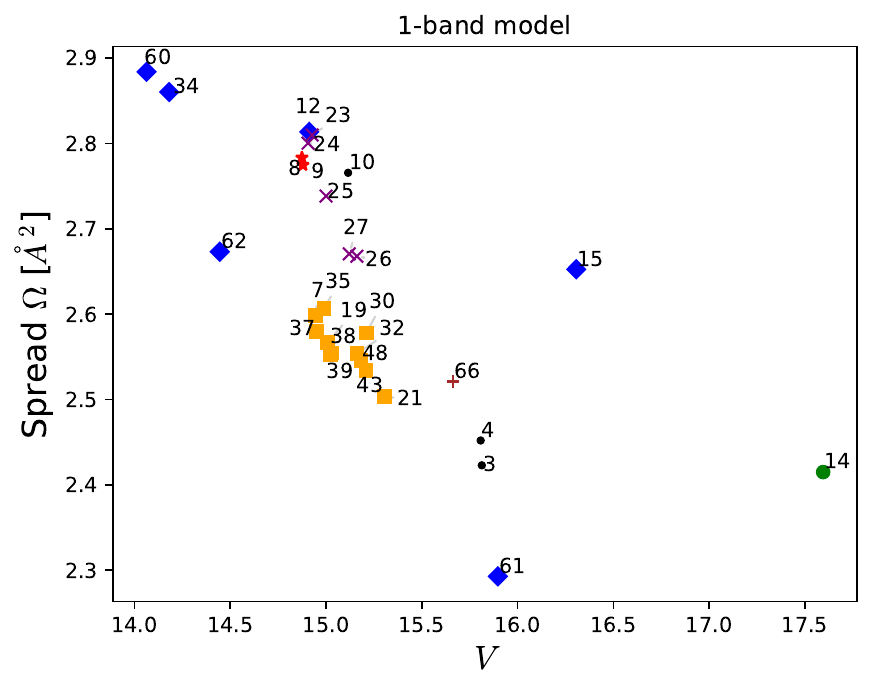}
\includegraphics[width=\columnwidth,page=10]{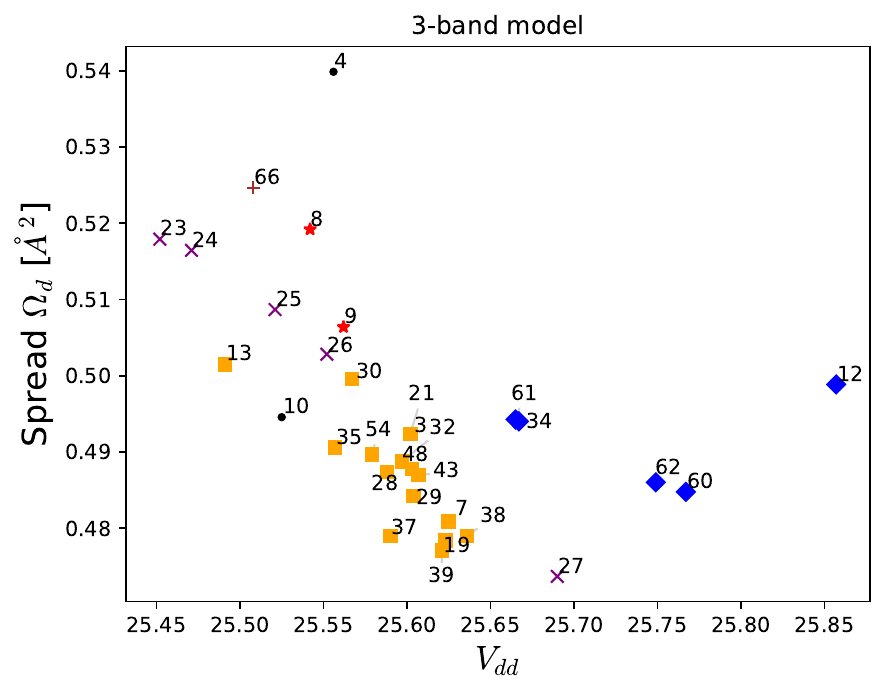}
\includegraphics[width=\columnwidth,page=11]{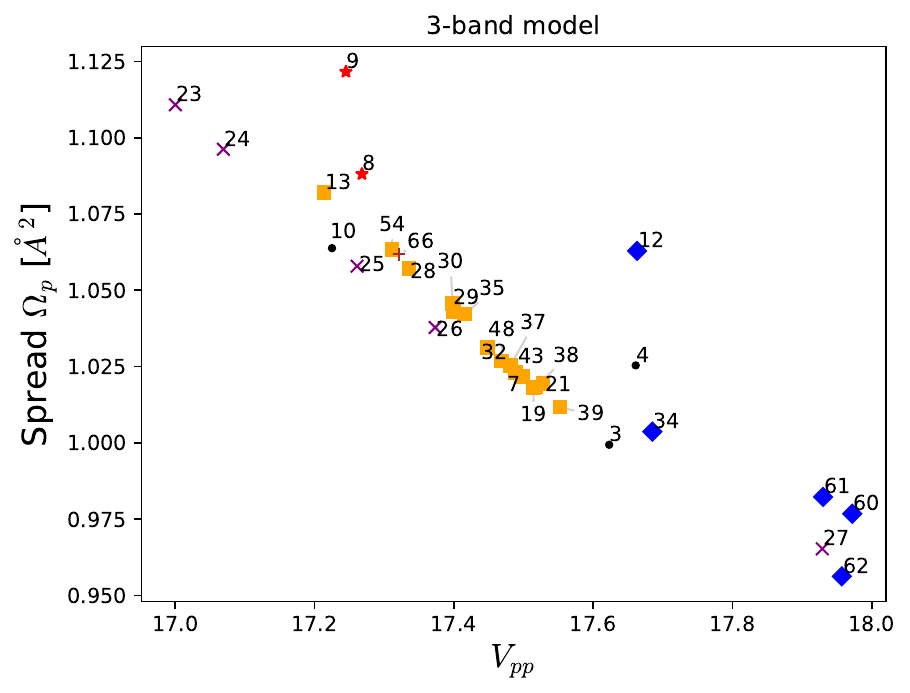}
\caption{Spreads of the MLWF vs. the corresponding on-site Coulomb interactions, showing anti-correlation, as expected. Spreads in the 1-band model are bigger than those in the 3-band model. In the 3-band model, the spreads of the oxygen-site orbitals are bigger than the spreads of the copper-site orbitals.
}
\label{fig:spreads}
\end{figure}

As already stated, the wannierization and the subsequent cRPA are done for all layers simultaneously - the number of wannier orbitals is $N_\mathrm{layers}\times N_\mathrm{bands\;in\;the\;model}$.
We have performed the wannierization using Wannier90 and RESPACK and confirm that the results are the same (by comparing the downfolded band-structures), whenever both algorithms converge the MLWF's.
The obtained spreads of the Wannier orbitals are presented in the Fig.~\ref{fig:spreads}. As expected, the spreads of MLWF's in the 1-band model are bigger than in the 3-band model, and in the 3-band model, oxygen-site orbitals are more spread out. In all cases, the spreads are found to anti-correlate with the bare Coulomb interactions, but especially so in the case of oxygen-site orbitals in the 3-band model.
Ultimately, we extracted the TB-parameters from the output of Wannier90 in an automated way, using a dedicated script.
The cRPA calculations were also performed starting from Wannier90 results.

\section{Parameters under consideration}
\label{app:parameters}
Here we list the parameters we have considered in our statistical analyses throughout Sections \ref{sec:relevant_examples} and \ref{sec:HubbardHolstein}.
\begin{itemize}
\item 1-band model 
    \begin{itemize}
     \item the nearest neighbor hopping $t$
     \item the next-n.n. hopping $t'$
     \item the third neighbor hopping $t''$,
     \item the bare Coulomb interaction $V$ and the cRPA-renormalized interaction at zero Matsubara frequency and at Matsubara frequency equal around 5eV, ${\cal U}(0)$ and ${\cal U}(5)$, respectively
     \item the effective n.n. antiferromagnetic interaction, at infinite Matsubara frequency, 5eV and 0eV: $J=t^2/V$, $J(5)=t^2/{\cal U}(5)$, $J(0)=t^2/{\cal U}(0)$
     \item the effective next-n.n. antiferromagnetic interaction, at infinite Matsubara frequency, 5eV and 0eV: $J'={t'}^2/V$, $J'(5)={t'}^2/{\cal U}(5)$, $J'(0)={t'}^2/{\cal U}(0)$
     \item the effective third-neighbor antiferromagnetic interaction, at infinite Matsubara frequency, 5eV and 0eV: $J''={t''}^2/V$, $J''(5)={t''}^2/{\cal U}(5)$, $J''(0)={t''}^2/{\cal U}(0)$
     \item the hopping ratios $t'/t$
     \item the n.n. vs. next-n.n. AFM interaction ratio $J/J'=t^2/{t'}^2$
     \item the interaction vs. hopping ratios ${\cal U}(0)/t$, ${\cal U}(5)/t$ and $V/t$
     \end{itemize}
\item 3-band model
    \begin{itemize}
     \item the bare charge-transfer gap $\varepsilon_d-\varepsilon_p$
     \item the hopping between the neighboring copper and oxygen $t_{pd}$
     \item the hopping between the nearest oxygens $t_{pp}$
     \item the hopping between the next-nearest oxygens $t'_{pp}$
     \item the local interaction on the copper site: the bare Coulomb $V_{dd}$, and the cRPA-renormalized values at zero and 5eV Matsubara frequency ${\cal U}_{dd}(0)$, ${\cal U}_{dd}(5)$
     \item the local interaction on the oxygen sites: the bare Coulomb $V_{pp}$, and the cRPA-renormalized values at zero and 5eV Matsubara frequency ${\cal U}_{pp}(0)$, ${\cal U}_{pp}(5)$
     \item the density-density interaction between copper and oxygen sites: the bare Coulomb $V_{dp}$, and the cRPA-renormalized values at zero and 5eV Matsubara frequency ${\cal U}_{dp}(0)$, ${\cal U}_{dp}(5)$
     \item the various ratios: $t_{pd}^2/{\cal U}_{dd}(0)$, $t_{pd}^2/{\cal U}_{dd}(5)$, $t_{pd}^2/V_{dd}$, $t_{pp}^2/{\cal U}_{pp}(0)$, $t_{pp}^2/{\cal U}_{pp}(5)$, $t_{pp}^2/V_{pp}$, ${\cal U}_{dd}(0)/t_{pd}$, ${\cal U}_{dd}(5)/t_{pd}$, $V_{dd}/t_{pd}$, ${\cal U}_{pp}(0)/t_{pp}$, ${\cal U}_{pp}(5)/t_{pp}$, $V_{pp}/t_{pp}$    
    \end{itemize}
\item Hubbard-Holstein and Emery-Holstein

Same as above, except that we replace all the parameters involving interactions at specific (finite) Matsubara frequencies (namely, 0 and 5eV) with the parameters of the Holstein boson and the corresponding coupling $g$, $A=g^2$ and $E$, for each of the orbitals in the model separately.
\end{itemize}

\section{cRPA}
\label{app:cRPA}
In this appendix, we briefly describe the cRPA implementation used in the present work. Additional implementation details can be found in Ref.~\cite{THCGW_Yeh2024}. 

The cRPA interaction $\mathcal{U}$ for a target subspace $\mathcal{C}$ is obtained by excluding screening processes internal to $\mathcal{C}$ from the full RPA polarization. In CoQuí, this is done on the imaginary-time axis by defining the constrained polarization
\begin{align}
    \Pi^{\boldsymbol{q}}_{\mathrm{cRPA}}(\boldsymbol{r}, \boldsymbol{r}'; \tau)
    =
    \Pi^{\boldsymbol{q}}_{\mathrm{RPA}}(\boldsymbol{r},\boldsymbol{r}';\tau)
    -
    \Pi^{\boldsymbol{q}}_{\mathcal{C}}(\boldsymbol{r}, \boldsymbol{r}'; \tau),
\end{align}
where $\Pi^{\boldsymbol{q}}_{\mathrm{RPA}}$ is the standard non-interacting RPA polarizability and $\Pi^{\boldsymbol{q}}_{\mathcal{C}}$ contains only particle-hole transitions within the target subspace. The latter is evaluated from the same non-interacting bubble diagram,
\begin{align}
    \Pi^{\boldsymbol{q}}_{\mathcal{C}}(\boldsymbol{r}, \boldsymbol{r}'; \tau)
    =
    \frac{1}{N_k}\sum_{\boldsymbol{k}}
    G_{\mathcal{C}}^{\boldsymbol{k}}(\boldsymbol{r},\boldsymbol{r}';\tau)
    G_{\mathcal{C}}^{\boldsymbol{k}+\boldsymbol{q}}(\boldsymbol{r}',\boldsymbol{r};-\tau),
\end{align}
but with the one-particle Green's function projected onto $\mathcal{C}$. Here $N_k$ is the number of sampled crystal momenta. 

For isolated target bands, the definition of $\Pi_{\mathcal{C}}$ is straightforward: one removes the particle-hole transitions within the corresponding band manifold. In the present work, however, the effective model subspaces are constructed from MLWFs obtained from highly entangled energy windows. In this situation, a direct projector-based construction of $G_{\mathcal{C}}$ may lead to over-screened or even negative static cRPA interactions, a known difficulty of cRPA for strongly entangled subspaces.

To avoid this pathology, we define $G_{\mathcal{C}}$ using a compact KS manifold selected at each $\boldsymbol{k}$ point. For a target subspace spanned by $N_{\mathcal{C}}$ MLWFs, we select the $N_{\mathcal{C}}$ KS states with the largest total projection weights onto $\mathcal{C}$. The projected Green's function $G^{\boldsymbol{k}}_{\mathcal{C}}$ is then formed within this selected KS manifold. This prescription removes the particle-hole transitions associated with the KS states that have the strongest MLWF character.
Empirically, across all cuprates considered in the present work, this choice yields non-negative static cRPA interactions and smooth frequency dependence on the imaginary axis without any system-dependent fine tuning, providing an automatic prescription for cRPA calculations in highly entangled subspaces.

We emphasize that downfolding was performed for all the layers simultaneously, in both wannierization and cRPA steps. 
In principle, cRPA produces results for any non-local density-density interactions, including those between the CuO$_2$ layers.
However, 
we find that those interactions are small (for example, at zero frequency, interlayer interactions are of order 0.5eV in multilayer compounds, and certainly much smaller in single-layer compounds), and are thus likely comparable to their systematic uncertainty. For this reason, we avoid looking at correlations between $T_c$ and interlayer coupling constants, and rather treat crystallographically inequivalent layers as separate data points. Having this in mind, one might be tempted to downfold different CuO$_2$ layers separately. Indeed, wannierization of individual layers gives similar results to what we have(see Supplemental Material\cite{SMrepo}), but the cRPA step based on such single-layer wannierizations necessarily gives unreasonable results (in multilayer cuprates).
The reason is that all CuO$_2$ layers contribute states near the Fermi level: If cRPA is done for the active subspace of a single CuO$_2$ layer, the screening due to other layers is very strong, and the resulting interactions are found to be very small. Such results are unlikely to be physically meaningful, especially because propagation through the active subspace of CuO$_2$ layers is expected to be strongly modified by interactions, and it cannot be properly captured by cRPA contributions. When downfolding all the layers simultaneously, none of the screening processes propagate through active subspaces of any CuO$_2$ layers, and no similar issue arises.

Finally, we have checked that the results were well converged with respect to numerical parameters such as the artificial temperature and the number of bands used in the cRPA calculation.

\begin{figure*}[ht!]
\centering
\includegraphics[width=0.45\textwidth,page=6]{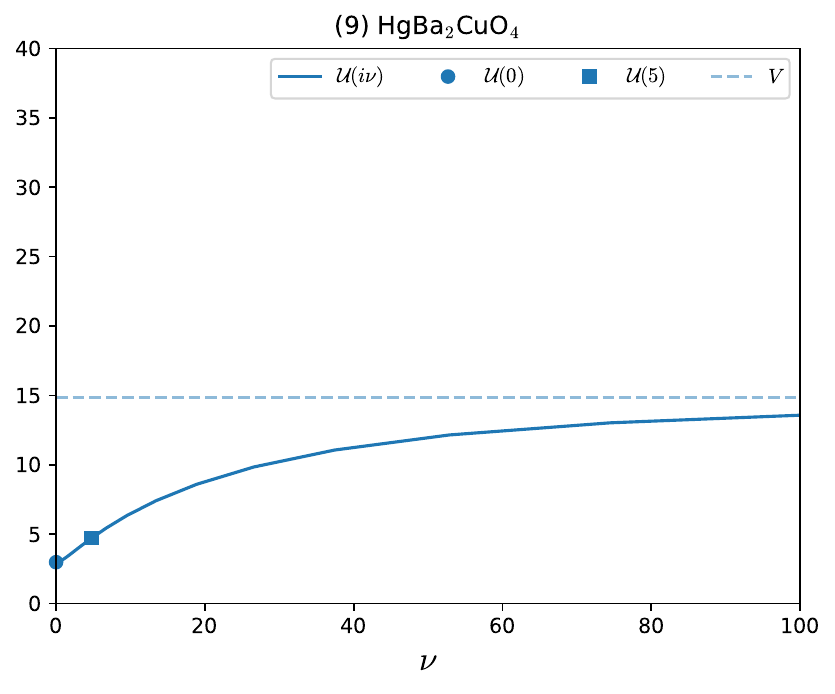}
\includegraphics[width=0.45\textwidth,page=6]{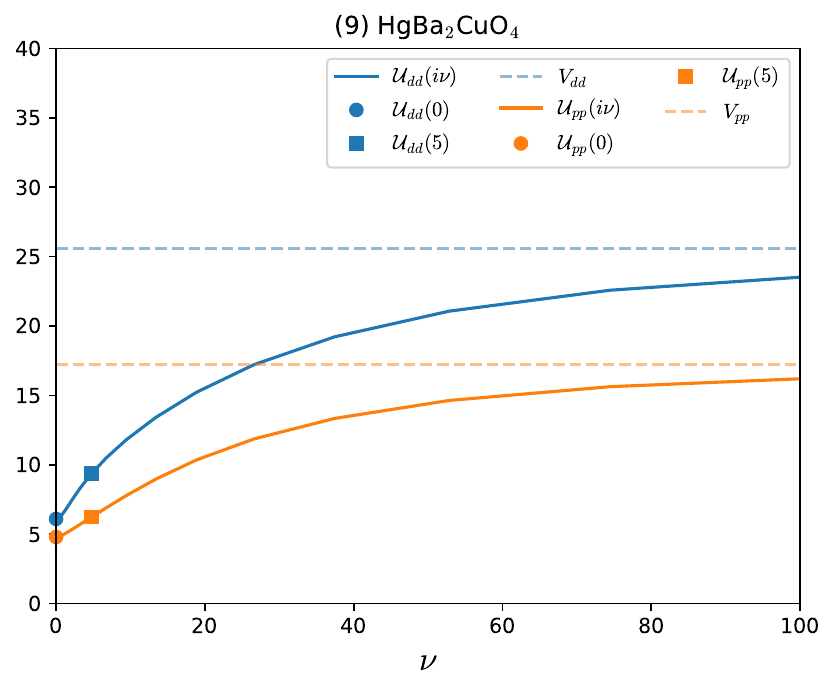}
\includegraphics[width=0.45\textwidth,page=22]{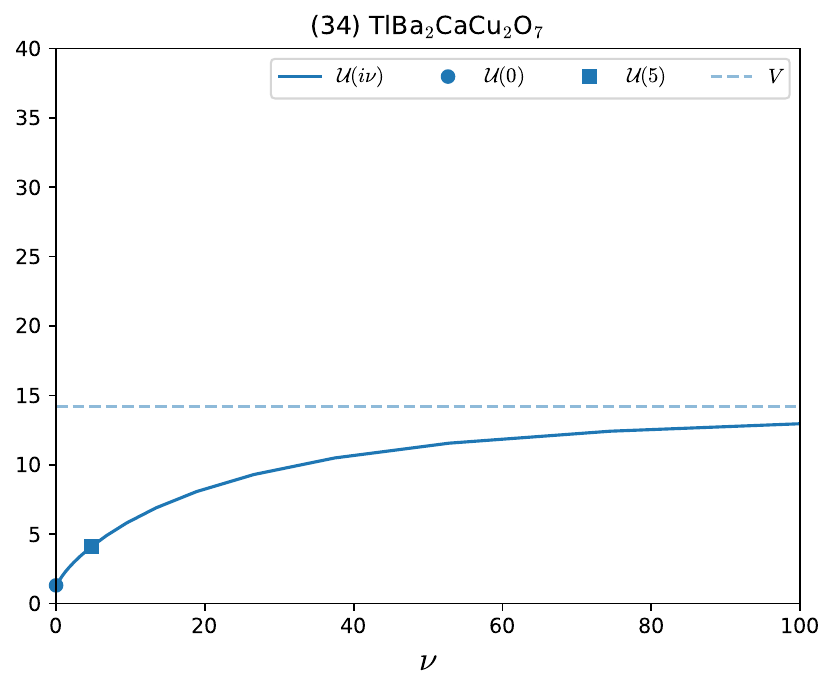}
\includegraphics[width=0.45\textwidth,page=23]{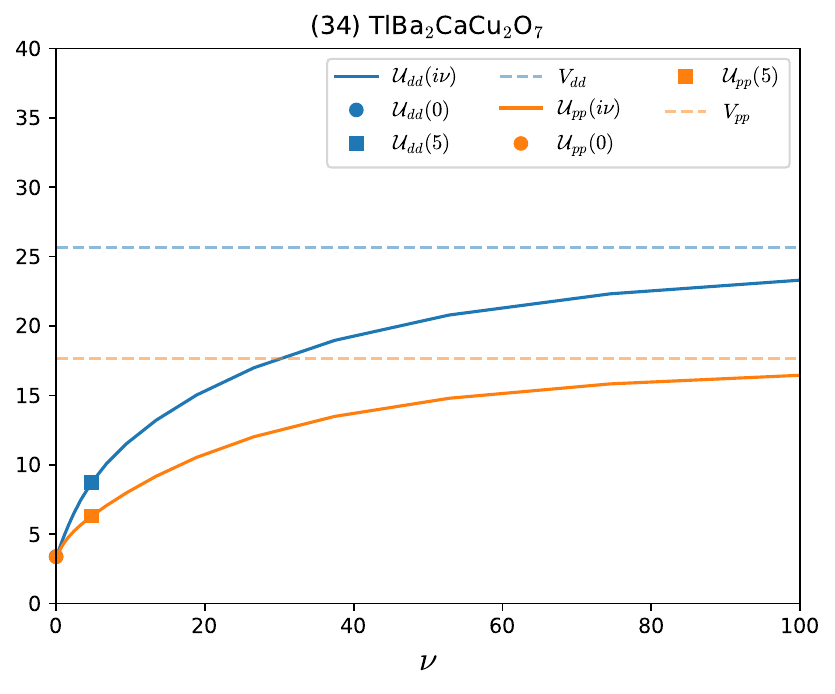}
\caption{Examples of cRPA results. Both axes are given in units of eV. Square, dot and the horizontal dashed line denote the values used in statistical analyses, the latter being the bare Coulomb interaction, which can be computed independently of the cRPA step. Left: 1-band model; right: 3-band model. Top right: more standard case when ${\cal U}_{dd}(0)>{\cal U}_{pp}(0)$; bottom right: surprising, but relatively common case when ${\cal U}_{dd}(0)<{\cal U}_{pp}(0)$. Universally, we find $V_{dd}>V_{pp}$. Bottom left: relatively common case when in the single-band model ${\cal U}(0)\sim 1.3$eV. 
}
\label{fig:cRPA_examples}
\end{figure*}

We show examples of our cRPA results in Fig.~\ref{fig:cRPA_examples}.

\begin{table} 
\caption{ Comparison of the bare Coulomb interaction $V$ and the static cRPA screened interaction ${\cal U}(0)$ for several cuprate compounds. All values are given in eV. Literature values are taken from Refs.~\onlinecite{Nilsson2019,Jang2016}. 
} 
\label{tab:cuprate_crpa_comparison} 
\begin{ruledtabular} 
\begin{tabular}{lcccccc} 
Compound 
& \multicolumn{2}{c}{This work} 
& \multicolumn{2}{c}{Ref.~\onlinecite{Nilsson2019}} 
& \multicolumn{2}{c}{Ref.~\onlinecite{Jang2016}} \\ \cline{2-3} \cline{4-5} \cline{6-7} 
& $V$ & ${\cal U}(0)$ 
& $V$ & ${\cal U}(0)$ 
& $V$ & ${\cal U}(0)$ \\ 
\hline La$_2$CuO$_4$ & 16.31 & 2.64 & 16.6 & 3.62 & -- & 3.15 \\ 
YBa$_2$Cu$_3$O$_6$ & 15.66 & 2.51 & 14.1 & 2.81 & -- & -- \\
HgBa$_2$CuO$_4$ & 14.87 & 2.98 & 15.5 & 3.42 & -- & 2.15 
\end{tabular}
\end{ruledtabular}
\end{table}

Table~\ref{tab:cuprate_crpa_comparison} compares the bare interaction $V$ and the static screened interaction ${\cal U}(0)$ obtained in this work with previous calculations~\cite{Jang2016,Nilsson2019} for several cuprate parent compounds. 
The underlying DFT electronic structures are expected to be broadly similar, but the details of the DFT approximation do differ between our work and those references (choice of exchange-correlation potential, pseudopotentials, treatment of spin polarization, etc.).
In addition, the parameters of wannierization are also different (outer and inner energy windows, etc.). 
We find that the bare interaction $V$ is dependent on those choices.
Furthermore, at the level of cRPA, the resulting ${\cal U}(0)$ depends on how the entanglement between the target low-energy subspace and the remaining Kohn--Sham bands is treated. In particular, Ref.~\onlinecite{Nilsson2019} employs a disentanglement procedure~\cite{Miyake2009}, whereas Ref.~\onlinecite{Jang2016} uses a weighting method~\cite{Ersoy2011} to define the constrained polarization.
Therefore, disagreement in the bare interaction value $V$ reflects differences in the spatial extent and shape of the MLWFs, while ${\cal U}(0)$ values additionally depend on how screening channels associated with the target subspace are excluded, as well as details of Kohn-Sham Hamiltonian. 
We emphasize that there is an important distinction between the uncertainty in MLWF's on the one hand, and the uncertainty in cRPA results on the other. The MLWF's define the physical meaning of the degrees of freedom (i.e. the orbitals) kept in the model. The choice of the model is in principle arbitrary and can only be proven appropriate in practice, e.g. by comparing experimental results with theory based on the given model. However, once the orbitals are chosen, the effective interactions are uniquely defined\cite{Profe2026} and can be understood as arising from a rigorous procedure to integrate out the degrees of freedom outside the model. In that sense, cRPA is an approximation scheme, and any uncertainty in its results reflect the fact that cRPA as a method is not uniquely defined, even if it is converged well with respect to numerical parameters (such as space-time discretization, number of DFT bands included, etc.). It is also based on the \emph{ab initio} part of the theory, i.e. it does depend on the choices made in constructing the Kohn-Sham Hamiltonian.

The comparison between our results and the two studies (Refs.~\onlinecite{Jang2016} and \onlinecite{Nilsson2019}) therefore provides a useful reference for assessing the level of variation associated with different Wannier constructions and constrained-screening prescriptions. As shown in Table~\ref{tab:cuprate_crpa_comparison}, the variation in ${\cal U}(0)$ between Refs.~\onlinecite{Nilsson2019} and~\onlinecite{Jang2016} is comparable to that observed between the present results and those previous calculations. Overall, the different $V$ and ${\cal U}(0)$ estimates are reasonably close.

However, the uncertainty of coupling constants (due to ambiguities of the procedure to derive lattice models) is comparable to the differences we observe between individual compounds. This is not necessarily a problem for our statistical analyses, because we apply the same procedure to all the compounds - all the biases of our approximation should hold roughly uniformly for all the compounds. Even if our method tends to, say, underestimate the coupling constants, it is still likely to capture the differences between the compounds the correct way.
Still, we observe that even the relative ordering of ${\cal U}(0)$ in different compounds shown in Table~\ref{tab:cuprate_crpa_comparison} is not the same in our work as in Refs.~\onlinecite{Jang2016} and \onlinecite{Nilsson2019}. In fact, already the bare Coulomb coupling $V$ is not in the same relative order. This reflects a substantial difference in the choice of Wannier orbitals and the inherent ambiguity in the approximation scheme employed in downfolding: the physical meaning of the models discussed in those references is slightly different. At present, a reasonable criterion for choosing the approximation is that it is well defined and robust enough to be applied to many materials, systematically, without the need for compound-specific tuning or human intervention - only that way one can hope to capture compound-dependent trends. Our procedure indeed proves to be robust across a large number of compounds and yields reasonable physical results consistently. Conversely, the procedures employed in Refs.~\onlinecite{Jang2016} and \onlinecite{Nilsson2019} have been tested on a relatively smaller number of compounds.

Finally, the unexpected finding ${\cal U}_{pp}(0)/{\cal U}_{dd}(0) \gtrsim 1$ we observe in some compounds might be an artifact of the cRPA approximation. Indeed, the bias of cRPA is likely orbitally dependent~\cite{Miyake2009,Kaltak2025_cRPA_spectral} in such a way that ${\cal U}_{dd}(0)$ is generically more underestimated than ${\cal U}_{pp}(0)$. The Cu $d_{x^{2}-y^{2}}$ orbital dominates the low-energy anti-bonding bands near the Fermi level in the DFT band structure and is therefore more sensitive to any residual low-energy screening that remains after the constrained polarization is constructed. This effect is potentially further enhanced by the fact that standard DFT predicts a metallic state for cuprate parent compounds, which might be artificially increasing the available screening channels between the target and rest spaces near the Fermi level. Nevertheless, even if the ratio ${\cal U}_{pp}(0)/{\cal U}_{dd}(0)$ is not well captured by cRPA, this is not of great importance for our statistical analysis: we focus on the compound-dependence of ${\cal U}_{pp}(0)$ and ${\cal U}_{dd}(0)$ independently, and we do not look at correlations between $T_c$ and ${\cal U}_{pp}(0)/{\cal U}_{dd}(0)$.

\section{Orientation in the parameter space}
\label{app:orientation}

\begin{figure*}[ht!]
\centering
\includegraphics[width=0.45\textwidth,page=1]{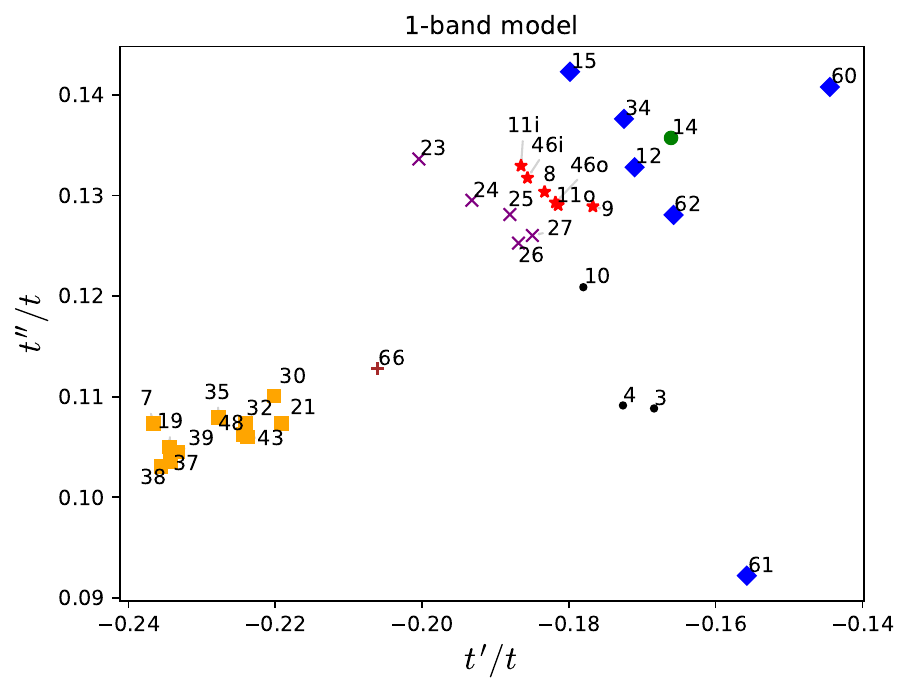}
\includegraphics[width=0.45\textwidth,page=3]{Figure_Systematic.pdf}
\caption{Region of the 1-band model parameter-space populated by our compounds. On the right panel, the axes are given in units of eV.  Clearly, different categories of our compounds populate slightly different parts of the phase diagram. It is surprising that the compounds with the highest and the lowest $T_c$'s  (red stars and purple crosses) are the most similar, and roughly in the center of the distribution. Also we observe that compound 60 is an extreme case where $|t''|\approx|t'|$. 
}
\label{fig:orientation_1band}
\end{figure*}

\begin{figure*}[ht!]
\centering
\includegraphics[width=0.45\textwidth,page=1]{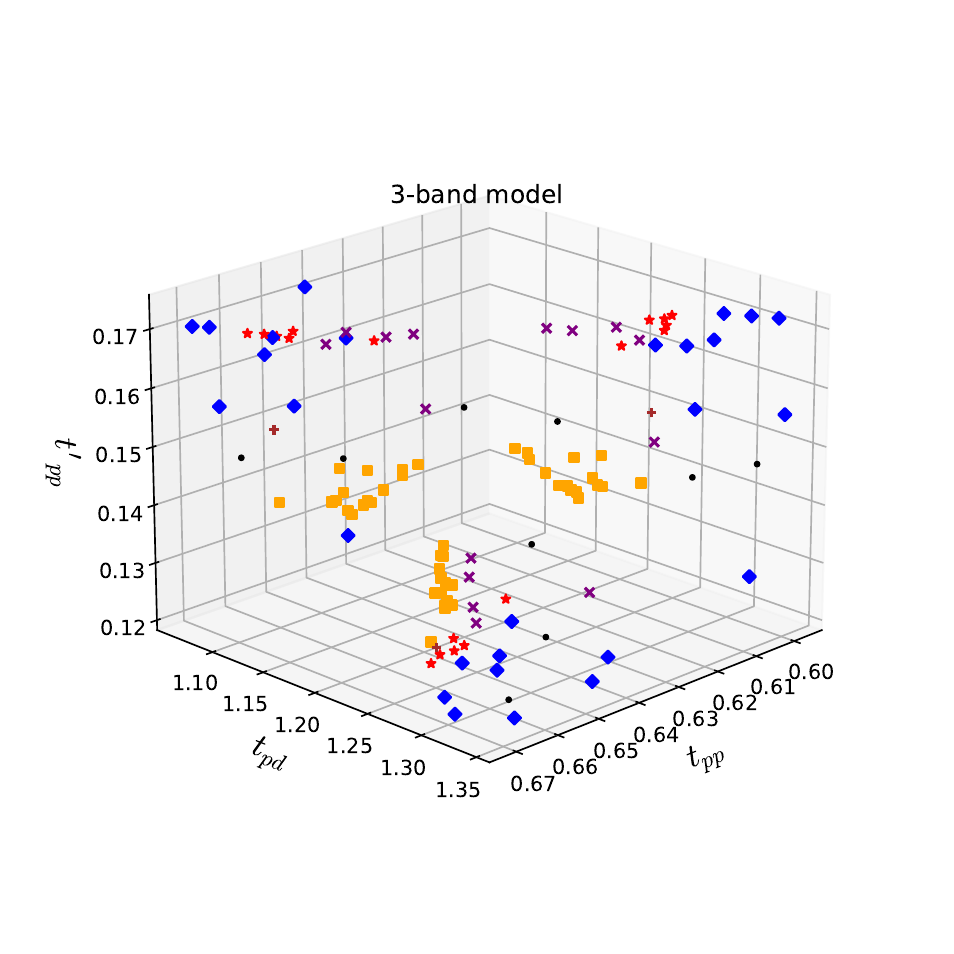}
\includegraphics[width=0.45\textwidth,page=2]{Figure_Systematic_3D.pdf}
\includegraphics[width=0.45\textwidth,page=16]{Figure_Systematic.pdf}
\includegraphics[width=0.45\textwidth,page=17]{Figure_Systematic.pdf}
\caption{Region of the 3-band model parameter-space populated by our compounds. Axes are given in units of eV. Similarly to the 1-band model, there is clear clustering of different categories of our compounds, and $T'$-structure $f$-element ternary compounds that have the lowest $T_c$'s (purple crosses) are close to the Ba/Hg compounds that have the highest $T_c$'s (red stars). The bare Coulomb interactions on the copper-site and oxygen-site orbitals correlate strongly. In absolute terms, the charge transfer gap $\varepsilon_d-\varepsilon_p$ varies the most of all tight-binding parameters. Bottom panels show parameters for the Holstein Hamiltonian terms to capture the effective retardation of interactions due to screening.
}
\label{fig:orientation_3band}
\end{figure*}

\begin{figure}[ht!]
\centering
\includegraphics[width=0.95\columnwidth,page=3, trim=0 2cm 0 2cm, clip]{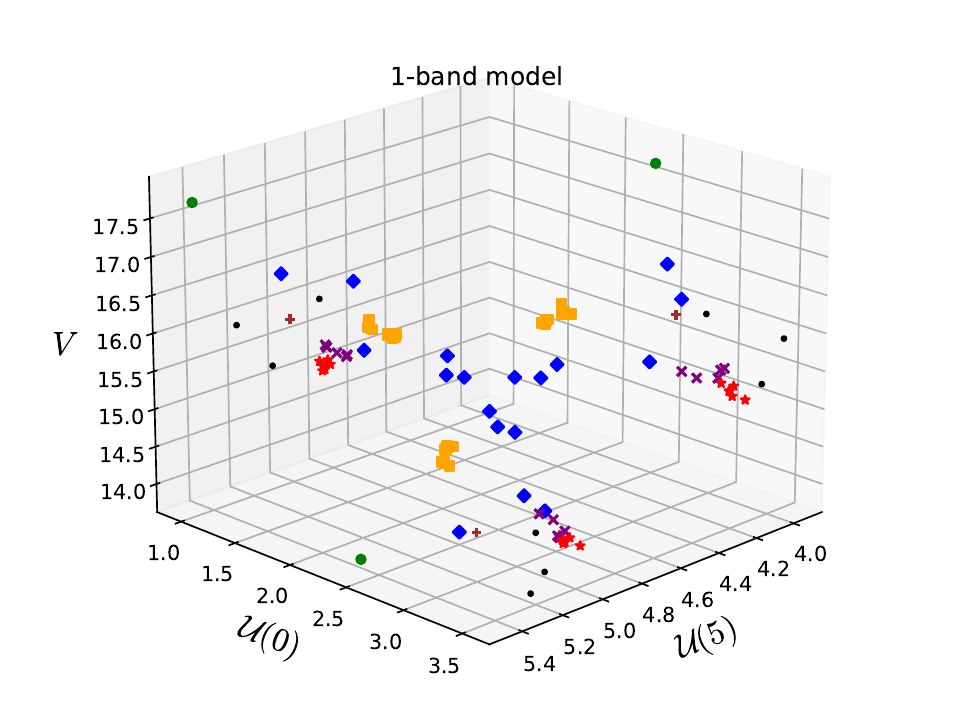}
\includegraphics[width=0.95\columnwidth,page=4, trim=0 2cm 0 2cm, clip]{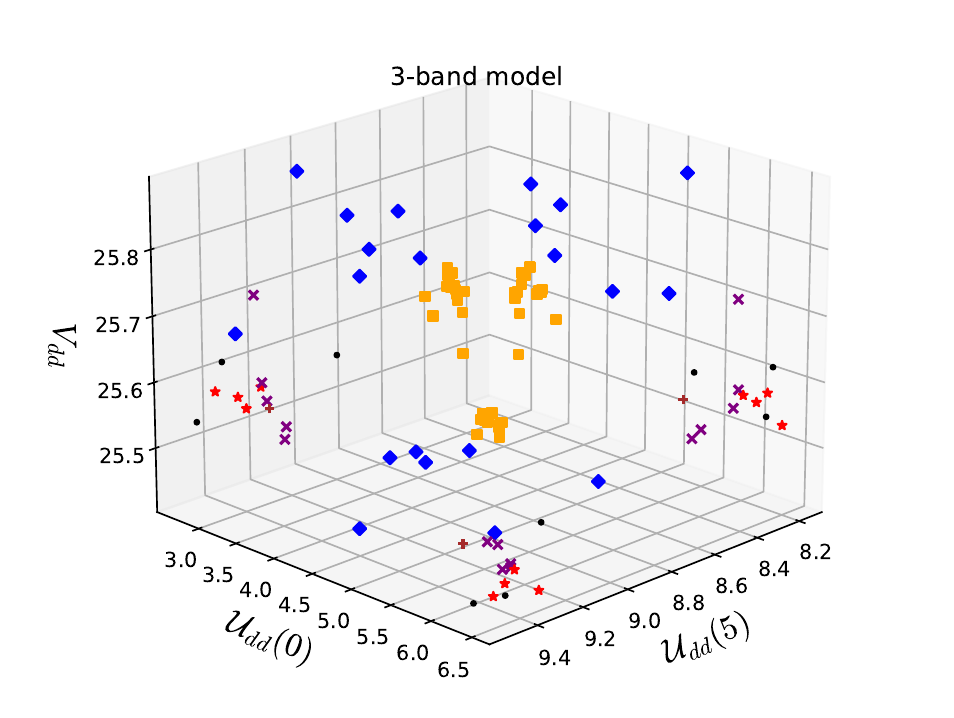}
\includegraphics[width=0.95\columnwidth,page=5, trim=0 2cm 0 2cm, clip]{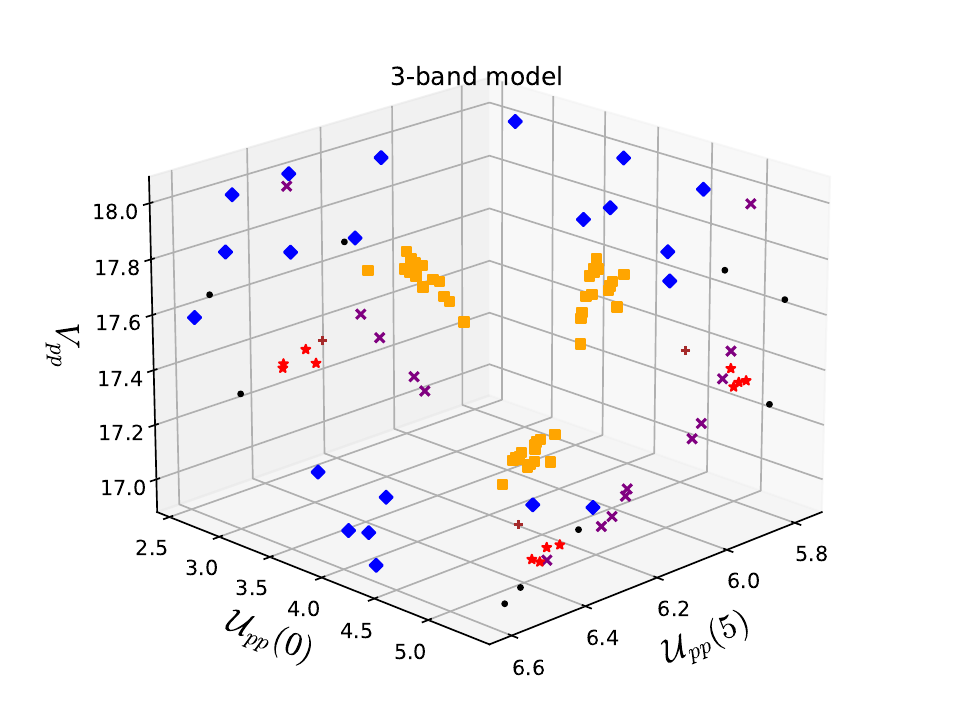}

\caption{Correlations between effective (cRPA) interactions at different Matsubara frequencies. Axes are in the units of eV.
}
\label{fig:Uw_correlations}
\end{figure}

It is of interest to inspect which parts of the parameter spaces of both models the cuprate compounds are populating.
We summarize this in Figs.~\ref{fig:orientation_1band} and \ref{fig:orientation_3band}. See captions for details.
Furthermore we analyze the correlations between cRPA interactions at different Matsubara frequencies, and summarize the results on Fig.~\ref{fig:Uw_correlations}.
As expected, there is some correlation between the values at different frequencies, but they are not particularly strong, especially so in the 3-band Emery model.

%\bibliography{refs}
\bibliography{Crtical_Temp_Data_Table}

\end{document}